\documentclass[a4paper]{article}
\usepackage{makeidx}
\usepackage{url}  
\usepackage[utf8]{inputenc}  
\usepackage{latexsym}  
\usepackage{amsmath} 
\usepackage{amssymb} 
\usepackage{graphicx}
\usepackage{authblk}
\usepackage{verbatim}

\newtheorem{lemma}{Lemma} 
\newtheorem{theorem}{Theorem}

\begin{document}
\title{Linear time construction of compressed text indices in compact space\thanks{This work was partially supported by Academy of Finland under grant 250345 (CoECGR).}}
\author{Djamal Belazzougui}
\affil{Helsinki Institute for Information Technology (HIIT),
Department of Computer Science, University of Helsinki, Finland.}

\maketitle
\begin{abstract}
We show that the compressed suffix array and the compressed suffix tree for a string of length $n$ over 
an integer alphabet of size $\sigma\leq n$ can both be built in $O(n)$ (randomized) 
time  using only $O(n\log\sigma)$ bits of working space. The previously fastest construction algorithms 
that used $O(n\log\sigma)$ bits of space took times $O(n\log\log\sigma)$ and $O(n\log^{\epsilon}n)$
respectively (where $\epsilon$ is any positive constant smaller than $1$). 
In the passing, we show that the Burrows-Wheeler transform of a string of length $n$ over 
an alphabet of size $\sigma$ can be built in deterministic $O(n)$ time and space 
$O(n\log\sigma)$. 
We also show that within the same time and space, we can carry many sequence analysis tasks
and construct some variants of the compressed suffix array and compressed suffix tree.

\end{abstract}
\section{Introduction}
The suffix tree~\cite{Wei73} is probably the most important text indexing data structure as it can be used for solving many string processing problems~\cite{Ap85,Gu97}. The suffix array~\cite{MM93} is another very popular data structure used for text indexing. 
Although it can not be used to solve as many problems as the suffix tree, its main advantage is the smaller constant in its space usage. 
Given a text of length $n$ over an alphabet of 
size $\sigma$, a suffix tree occupies $O(n\log n)$ bits of space while a suffix array occupy $n\log n$
~\footnote{In this paper $\log n$ stands for $\log_2 n$.} bits. 

The last decade has witnessed the rise of space efficient versions of the suffix array~\cite{GV05,FM05} and the suffix tree~\cite{Sa07a}. In contrast to their non compressed versions they occupy only $O(n\log\sigma)$ bits of space, which saves a factor $\Theta(\log_\sigma n)$ and is only a constant factor larger than the original text (which occupies $n\log\sigma$ bits). 
Any operation that can be implemented on a suffix tree can also be implemented on the compressed suffix tree (henceforth \textsc{CST}) at the price of a slowdown that can vary from $O(1)$ to $O(\log^\epsilon n)$ time. Thus any algorithm or data structure that uses the suffix tree can also be implemented using the \textsc{CST} with a slowdown at most $O(\log^\epsilon n)$. 
%Since their inceptions, the area of compressed text indexing has become very active~\cite{NM07} and most of the  
%the basic problems in the area are now well understood. Among the remaining l

While a \textsc{CST} occupies a smaller space, when compared to its non-compressed counterpart, its construction suffers from a large slowdown if it is restricted to use a space that is only a constant factor away from the final space. More precisely a \textsc{CST} can be built in $O(n\log^\epsilon n)$ time (where $\epsilon$ is any constant such that $0<\epsilon<1$) using $O(n\log\sigma)$ bits~\cite{HSS09}. Alternatively it can be built in $O(n)$ time but using (non-succinct) $O(n\log n)$ bits of space if one uses an optimal suffix tree construction algorithm to build the (non-compressed) suffix tree~\cite{Fa97} and then compresses the representation.

%Other possible tradeoffs are a $O(n)$ time construction with $O(n\log^\epsilon n)$ bits of space or $O(n\log\log n)$ time using $O(n\log\sigma\log\log n)$ bits of space. 

%A recent result~\cite{NP07} shows that the Burrows-Wheeler transform 
%and the compresses suffix array can be built in time $O(n)$ while using $O(n\log\sigma(\log_\sigma n)^\alpha)$ bits of space, where $\alpha=\log_3 2$. A direct corollary of that algorithm is that the compressed suffix tree can be constructed in time $O(n)$
%using space only $O(n\log\sigma(\log_\sigma n)^\alpha)$. 
% A direct consequence 
%of the result is that the compressed suffix tree can be built in time $O(n)$
%using space $O(n\log\sigma(\log_\sigma n)^\epsilon)$, where $\epsilon$ is any 
%constant such that $\epsilon>1$ . 
The compressed version of the suffix array (the \textsc{CSA}) does not suffer the same slowdown as the compressed version of the suffix tree, since it can be built in time $O(n\log\log\sigma)$~\footnote{This bound should actually read as $O(n\cdot\mathtt{max}(1,\log\log\sigma))$. } when the space is restricted to be $O(n\log\sigma)$ bits~\cite{HSS09}. 
Alternatively it can be built in (deterministic) time $O(n)$ using $O(n\log\sigma \log\log n)$ bits of space~\cite{OS09}.
%In contrast the (non-compressed) suffix array can be built in $O(n)$ time, but requires $O(n\log n)$ bits of space~\cite{KA03,KSPP05,KSB06}. 

The main result of this paper is to show that both the \textsc{CST} and the \textsc{CSA} can be built in randomized linear time using $O(n\log\sigma)$ bits of space. In the passing, we show that the Burrows-Wheeler transform of a string of length $n$ over 
an alphabet of size $\sigma$ can be built in deterministic $O(n)$ time and space $O(n\log\sigma)$. We also 
show that many sequence analysis applications can be carried out within the same time and space bound.

We note that the (non-compressed) suffix array and suffix tree can be built in time deterministic $O(n)$ as opposed to the randomized $O(n)$ we achieve. The randomization is due to hashing. However, we also note that hashing is also needed to represent the non-compressed suffix tree if one wants to support the fundamental child operation in constant time~\footnote{The constant time child operation can be used to match a pattern of length $m$ against the suffix tree in time $O(m)$.}. In that case building the representation needs randomized $O(n)$ time. 

\begin{comment}
Otherwise, if one insists on deterministic linear time, then the child operation (on a non-compressed suffix tree) can only be supported in $O(\log\log\sigma)$ time.
If one abandons the requirement that certain operations on the compressed suffix tree be supported in constant time (but supported in $O(\log\log\sigma)$ time), then we show that the compressed suffix tree can be built in deterministic $O(n\log\log\sigma)$ time. 
\end{comment}
\section{Organization and overview of the techniques}
In~\cite{BGOS11,BGOS13} a technique was introduced which allows to enumerate 
all the suffix tree nodes (actually the corresponding suffix array intervals) 
using space $n\log\sigma+O(n)$ bits and $O(n\log\sigma)$ time,
based solely on the Burrows-Wheeler transform and $O(n)$ bits of extra-space 
used to store a succinctly represented queue and a bitvector. 
It was predicted that the method would allow to solve many problems 
that relied on the \textsc{CST} in $O(n\log\sigma)$ time instead of $O(n\log^\epsilon n)$
time. In~\cite{BBO12} the method was successfully applied to the maximal repeat problem. 
In~\cite{BCKM13}, many more applications were described and 
a new enumeration technique based on the bidirectional Burrows-Wheeler transform
was introduced. 
The new technique allows to enumerate intervals in constant time per interval allowing to solve 
many problems in $O(n)$ time, once the required data structure were built. 
However, no efficient method to build those data structures was described. 

One of the contributions of this paper is a third enumeration technique that is more 
space-efficient than the 
previous ones and that might be of independent interest. 
It is a modification of the technique of~\cite{BGOS13} 
but uses a stack instead of a queue and eliminates the need for
a bitvector. 

The \textsc{CST} has three main components, the Burrows-Wheeler 
transform, the tree topology and the permuted longest common prefix 
array. 

We will show that the enumeration 
technique allows to easily build the \textsc{CST} topology in asymptotically 
the same time needed to enumerate the intervals. 
We will also show how to efficiently build the permuted longest common prefix array 
based on the bidirectional Burrows-Wheeler. The technique might be of independent 
interest and might prove useful to solve other kinds of problems of the same flavor. 

Finally, we will show that a variation of our new enumeration technique 
allows to build the Burrows-Wheeler transform in deterministic linear time. 
For that, we will reuse the algorithm described in~\cite{HSS09}. 
That algorithm proceeds in $O(\log\log n)$ steps, where each step 
involves the merging of the Burrows-Wheeler transforms of two strings 
of geometrically increasing sizes with the last step involving 
two strings of length $n/2$ each. 
We show that each merging can be done in linear time, resulting in an 
overall linear time. 
%For lack of space we defer the description of some building blocks 
%to the full version of the article~\cite{Be14}. The omitted descriptions
%are often technical and do not involve any interesting new techniques. 

The paper is organized as follows: in section~\ref{sec:back_prel}
we introduce the model and assumptions
and the basic structures from the literature 
that will be used in our algorithms. 

In section~\ref{sec:small_alpha_build} we will describe the algorithms 
that build the suffix tree topology and the longest common prefix array. 
That shows that a basic \textsc{CST} can be built in time $O(n\log\sigma)$. 
In section~\ref{sec:big_alpha_build}, we present our new enumeration technique
and use it to build the Burrows-Wheeler transform in linear time. The Burrows-Wheeler 
transform is the main component in most \textsc{CSA} variants. In the full 
version we will show that the remaining components of the \textsc{CSA}
(at least the ones used in some recent variants of the CSA) and the \textsc{CST}
can be built in randomized linear time. 
We finally outline some applications of our results in section~\ref{sec:applications}.

\section{Background and preliminaries}
\label{sec:back_prel}
We assume the unit-cost RAM model with word size $w=\Theta(\log n)$ bits and with all usual arithmetic and logic
operations taking constant time (including multiplication). 
We assume an integer alphabet $[1..\sigma]$. Throughout the paper, we will assume that $\sigma\leq n^{1/3}$. Otherwise, there already exist efficient methods to build the suffix tree~\cite{Fa97} (and hence the \textsc{CST} too) and the suffix array~\cite{KA03,KSPP05,KSB06} (and hence the \textsc{CSA}) in $O(n)$ time using $O(n\log n)=O(n\log\sigma)$ bits of space. 
In section section~\ref{sec:text_indexes} we give a brief description of the suffix array and the suffix tree.
In section~\ref{sec:succ_DS}, we describe the succinct data structures used in this paper. Finally, in section~\ref{sec:compressed_text_idx}, we describe the compressed text indices used in this paper. 
We assume that the reader is familiar with the two major text indexing data structures, the suffix tree and the suffix array. If not, he can find a brief description in section~\ref{sec:text_indexes}. We also assume that the reader is familiar with standard succinct data structures, like rank and select data structures, succinct prefix-sum representations and wavelet trees. 
%For lack of space we describe those data structures in appendix~\ref{sec:succ_DS}
\subsection{Text indexes}
\label{sec:text_indexes}
\subsubsection{Suffix trees}
A suffix tree for a string $T[1..n-1]$ is a compacted trie built on top of all the suffixes of the string $T[1..n-1]\$$, where 
$\$$ is a character not present in $T[1..n-1]$ and lexicographically smaller than all characters that appear in $T[1..n-1]$. 
Every internal node in the suffix tree is labeled by a path $p$ that corresponds to a $right$ maximal-factor of $T$. 
A factor (substring of $T$) $p$ is said to be right-maximal if and only if we have at least two distinct characters $a$ and $b$ such that the two factors $pa$ and $pb$ appear in $T[1..n-1]\$$. 
Every leaf of the suffix tree is labeled by a suffix and stores a pointer to it. All the suffix tree leaves are sorted in 
left-to-right order according to the lexicographic order of their corresponding suffixes. 

Suppose that a node $x$ labeled by the path $p$ is a child of a node $y$ labeled by the path $p'=pcq$, where $c$ is a character and $q$ is a possibly empty string. Then the edge that connects $y$ to $x$ is labeled by the character $c$. 
Since the suffix tree is a compacted trie with no node of degree $1$ and with $n$ leaves, it can have at most $n-1$ internal nodes. 

\subsubsection{Suffix arrays}
A suffix array for a string $T[1..n-1]$ is an array $A[1..n]$ such that $A[1]=n$ and $A[i]=j$ for $i>1$ if and only if the suffix $T[j..n-1]$  is of rank $i$ among all suffixes of $T[1..n-1]$ sorted in lexicographic order. 
There exists a strong relationship between suffix trees and suffix arrays. More precisely, the suffix $A[i]$ is exactly the one that labels the $i$th suffix tree leaf in left-to-right order.

\subsubsection{Suffix array intervals}
Given any factor $p$ that appears in the text $T$, there exists a corresponding interval $[i..j]$ such that the subarray $A[i..j]$ contains the pointers to all the $j-i+1$ suffixes of the text that are prefixed by $p$. Given a factor $p$, its suffix array interval is the same as that of the string $p'$, where $p'$ is the shortest right-maximal string prefixed by $p$ if it exists, or the only suffix prefixed by $p$ if not. 

There is a bijection between suffix tree nodes and the suffix array intervals. Every suffix tree node uniquely corresponds to a suffix array interval and vice-versa. More precisely, the leaves under the suffix tree node labeled by a path $p$ are precisely all the leaves labeled by suffixes $A[i]$,$A[i+1]\ldots A[j]$, in left to right order, where $[i..j]$ is the suffix array interval that corresponds to the factor $p$. 
The bijection implies that the total number of suffix array intervals is at most $2n-1$. 

\subsubsection{Weiner and Suffix links}
A suffix tree can be augmented with Weiner and Suffix links. 
A suffix link is a directed edge that connects: 
\begin{enumerate}
\item A leaf corresponding to a suffix $cp$ to the leaf corresponding to the suffix $p$, where $c$ is a character. 
\item An internal node labeled by the right-maximal factor $cp$ to the internal node labeled by the factor $p$ (where $p$ is by necessity also right-maximal), where $c$ is a character. 
\end{enumerate}

An explicit Weiner link is a directed edge labeled by a character $c$ that connects a node $x$ to a node $y$ such that:
\begin{enumerate}
\item There exists a suffix link that connects $y$ to $x$. 
\item The right-maximal factor that labels $y$ is prefixed by character $c$. 
\end{enumerate}
In other words, an explicit Weiner link connects a node labeled by path $p$ to a node labeled by 
the path $cp$. 

An implicit Weiner link is a directed edge labeled by a character $c$ that connects the node $x$ to a node $y$ such that: 
\begin{enumerate}
\item The node $x$ is labeled by a path $p$. 
\item No node is labeled by $cp$. 
\item The node $y$ is labeled by $cp'$, where $cp'$ is the shortest string such that $p$ is a proper 
prefix of $p'$ and $cp'$ is the label of a node in the tree. 
\end{enumerate}
The total number of explicit Weiner link is linear since every node can only be the destination of one explicit Weiner link. 
It turns out that the number of implicit is also linear (see for example~\cite{BNtalg14} for a proof). 

\subsection{Succinct data structures}
~\label{sec:succ_DS}
\subsubsection{Rank and select}
Given an array $A[1..n]$ of $n$ elements from $[1..\sigma]$, we would wish to support the following three operations: 
\begin{enumerate}
\item $\mathtt{access}(i)$, return $A[i]$. 
\item $\mathtt{rank}_c(i)$, return the number of occurrences of character $c$ in $A[1..i]$. 
\item $\mathtt{select}_c(j)$, return the position of the occurrence number $j$ of character $c$ in $A$. That 
is, return the only position $i$ such that $\mathtt{rank}_c(i)=j$ and $A[i]=c$
\end{enumerate}
In~\cite{GMR06} it is shown how to build two different data structures that both occupy $n\log\sigma(1+o(1))$, 
but with different tradeoffs for $\mathtt{access}$, $\mathtt{select}$ and $\mathtt{rank}$ queries.

The first one supports $\mathtt{select}$ in constant time and $\mathtt{access}$, $\mathtt{rank}$ in $O(\log\log\sigma)$ time. 

The second one supports $\mathtt{access}$ in constant time, $\mathtt{select}$ in $O(\log\log\sigma)$ time 
and $\mathtt{rank}$ in $O(\log\log\sigma\log\log\log\sigma)$ time. 

In~\cite{GOR10} the time of $\mathtt{rank}$ of the second data structure was improved to
$O(\log\log\sigma)$ time while maintaining the same space bound. 
We note that for the special case $\sigma=2$ ($A$ is a bitvector) there exists older solutions which use $n+o(n)$ bits of space 
and support $\mathtt{rank}$ and $\mathtt{select}$ in constant time~\cite{Cl96,Mun96}. 

\subsubsection{Prefix-sum data structure}
Given an array $A[1..n]$ of numbers which sum up to $U$, a prefix-sum 
data structure is a structure which allows given any $i\in[1..n]$ 
to return the sum $\sum_{1\leq j\leq i}A[j]$. 
Using Elias-Fano~\cite{El74,Fa71} encoding in combination with bitvectors with constant time select support allows
to build in linear time a data structure which occupies $n(2+\log(U/n))+o(n)$ bits of space and that 
answers to prefix-sum queries in constant time.

\subsubsection{Wavelet trees}
The wavelet tree~\cite{GGV03} over a sequence of $n$ elements from $[1..\sigma]$ 
is a data structure which occupies $n\log\sigma+o(n)$ bits~\cite{GGGRS07} and supports 
$\mathtt{access}$, $\mathtt{rank}$ and $\mathtt{select}$ queries in $O(\log\sigma)$ time. 
Thus a wavelet tree is slower but more space-efficient 
than the structures of~\cite{GMR06}. It is also considerably simpler.

\subsubsection{Range minimum and range color reporting queries}

A range minimum query data structure (\textsc{rmq} for short)
is a data structure built on top of an integer 
array $C[1..n]$ and that is able to answer to the following queries: given 
a range $[i..j]$, return the index $x\in[i..j]$ such that the element 
$C[x]$ is the smallest among all the elements in $C[i..j]$ (ties are broken 
arbitrarily). There exists a range minimum data structure which occupies $2n+o(n)$
bits of space and which answers to a query in constant time, without 
accessing the original array~\cite{Fi10}.

Given an array $A[1..n]$ of $n$ elements from $[1..\sigma]$, we can build a data structure of size $2n+o(n)$ bits 
so that we can report all the $\mathtt{occ}$ distinct colors in an interval $A[i..j]$ in time $O(\mathtt{occ})$. 
The data structure is a \textsc{rmq} built on top of an array $C[1..n]$ where $C[i]=j$ 
if and only $j$ is the maximal index such that $A[j]=A[i]$ and $j<i$ (that 
is $C[i]$ stores the position of the previous occurrence of character $A[i]$). 
The algorithm for reporting the colors, needs 
to do $O(\mathtt{occ})$ accesses to the arrays $C$ and $A$. 
Such an algorithm was first described in~\cite{Mu02}. 
Subsequently, it was shown that the same result could be 
achieved by only doing $O(\mathtt{occ})$ accesses to $A$~\cite{Sa07b}.

\subsubsection{Succinct tree representations}
The topology of a tree of $n$ nodes can be represented using $2n+o(n)$ bits so that 
many operations can be supported in constant time~\cite{NStalg14}. Among them are basic 
navigation operations like going to a a child or to the parent of a node, but also more 
advanced operations like the $\mathtt{lca}$ which returns the 
the lowest common ancestor of two nodes, or the operations $\mathtt{leftmost\_leaf}$ and 
$\mathtt{rightmost\_leaf}$ which for a node $y$, return the indexes $i+1$ and $j+1$ 
of the leftmost and rightmost leaves $y$ and $z$
in the subtree of $x$, where $i$ and $j$ are respectively
the number of leaves of tree that lie on the left of $y$ and $z$. 
 
The topology of a tree over $t$ nodes can be described using a sequence 
of $2t$ balanced parenthesis built as follows: start 
with an empty sequence then write an opening parenthesis, 
recurse on every child of the root in left-to-right order 
and finally write a closing parenthesis. Another way to view 
the construction of the balanced parenthesis sequence is as 
follows: we do an Euler tour of the tree and write an opening 
parenthesis every time we go down and a closing parenthesis 
when we go up the tree. 

\subsubsection{Monotone minimal perfect hashing}
Given a set $S\subset[1..U]$ with $|S|=n$, a 
monotone minimal perfect hash function (henceforth $\textsc{mmphf}$)
is a function $f$ from $U$ into $[1..n]$ such that $f(x)<f(y)$
for every $x,y\in S$ with $x<y$. In other words 
if the set of keys $S$ is $x_1<x_2<\ldots <x_n$, 
then $f(x_i)=i$ (the function returns the rank 
of the key it takes as an argument).  
The function is allowed 
to return an arbitrary value on any $x\in U/S$. 

In~\cite{BBPV09}, it is shown that there exists a scheme which given 
any set $S\subset[1..U]$ with $|S|=n$, builds 
a $\textsc{mmphf}$ on $S$ that occupies $O(n\log\log(U/n))$
bits of space and such that $f(x)$ can be evaluated in constant time.

\subsection{Compressed text indexes}
\label{sec:compressed_text_idx}
\subsubsection{The Burrows-Wheeler transform}
Given a string $X$, the Burrows-Wheeler transform (henceforth \textsc{bwt}) is obtained as follows
%~\footnote{In fact the Burrows-Wheeler transform works on any string, but here we only def}: 
we sort all the $n$ rotations of $X$ and take the last character in each rotation in sorted order. 
There is a strong relation between the \textsc{bwt} of the string $T[1..n-1]\$$ over an alphabet of size $\sigma$ (where $\$$ is a character that does not appear in $T$) and the suffix array of the string $T[1..n-1]$. The former can be obtained from the latter just by setting $\textsc{bwt}[i]=T[A[i]-1]$ whenever $A[i]>1$ and $A[i]=\$$ otherwise. 
It is well-known that the \textsc{bwt} can be built in $O(n\log\log\sigma)$ time~\cite{HSS09}, while using $O(n\log\sigma)$ bits of (temporary working) space. 

\subsubsection{FM-index}
The FM-index~\cite{FM05} is a succinct text index built on top of the \textsc{bwt}. There are many variants of the FM-index, but they all share the same basic components: 
\begin{enumerate}
\item The \textsc{bwt} of the original text. 
\item The array $C[1..\sigma]$ which stores in $C[i]$, the number of occurrences of all characters $b<i$ in $T[1..n-1]\$$. 
\item A sampled suffix array $\mathtt{SSA}$ built as follows: given a sampling factor $b$, for $i=1,2\ldots n$ append to $\mathtt{SSA}$ the value $\mathtt{SA}[i]$ if and only if $\mathtt{SA}[i]\bmod b=1$ or $i=n$. 
\end{enumerate}
The search for a pattern $p$ in an FM-index amounts to finding in the \textsc{bwt} the interval of rotations prefixed 
by $p$. The sampled suffix array allows to report the starting position of every such rotation in $T[1..n-1]\$$ in time linear in $b$ (with possibly a small dependence on $\sigma$ or $n$). 
Finding the interval of rotations for a pattern $p$ of length $m$ is done backwards. First determine the interval of $p[m]$, then the interval of $p[m-1..m]$ and so on until we get to the interval of $p[1..m]=p$. 
The search in the FM-index is based on Weiner links which can be efficiently simulated if the \textsc{bwt} sequence is augmented so that it supports $\mathtt{rank}$ queries. More precisely, given the interval $[i_1,j_1]$ that corresponds to a factor $p$ and a character $c$, the interval $[i_2,j_2]$ that corresponds to the factor $cp$ can be computed as $i_2=\mathtt{rank}_c(i_1-1)+C[c]+1$  and 
$j_2=\mathtt{rank}_c(j_1)+C[c]$. If $i_2>j_1$, then it is deduced that there is no occurrence of the factor $cp$ in the string $T[1..n-1]\$$. The Weiner links were initially defined for suffix trees and operate on suffix tree nodes which translate to intervals of suffixes. Here we use Weiner links to operate on intervals of rotations of the \textsc{bwt}, since there is a bijection between suffixes and rotations (except for the rotation that starts with $\$$ which is irrelevant). Henceforth, we will no longer talk about rotations, but instead talk about suffixes. 

The intuition behind the Weiner link formulae is as follows: we are given the set of all suffixes that are prefixed by $p$ and we wish to compute the set of suffixes prefixed by $cp$. For that it suffices to note that all we need is to compute the suffixes that are prefixed by $p$ and preceded by $c$. By doing $\mathtt{rank}_c(i_1-1)$ we compute the number of suffixes that are prefixed by some  $p'$ lexicographically smaller than $p$ and preceded by $c$ (in the text). Equivalently this represents the number of suffixes prefixed by $cp'$, where $p'$ is lexicographically smaller than $p$. Thus the first suffix prefixed by $cp$ among all those prefixed by $c$ (if it exits) must be at position $\mathtt{rank}_c(i_1-1)+1$ which means that its lexicographic rank among all suffixes is $i_2=C[c]+\mathtt{rank}_c(i_1-1)+1$. Then $n_c=\mathtt{rank}_c(j_1)-\mathtt{rank}_c(i_1-1)$ represents the number of suffixes prefixed by $p$ and preceded by $c$ which is actually the number of suffixes prefixed by $cp$. Therefore, we deduce that $j_2=i_2+n_c-1=(C[c]+\mathtt{rank}_c(i_1-1)+1)+(\mathtt{rank}_c(j_1)-\mathtt{rank}_c(i_1-1))-1=\mathtt{rank}_c(j_1)+C[c]$. 

The time to compute a Weiner link is thus dominated by the time needed to do a $\mathtt{rank}$ query which is $O(\log\log\sigma)$ or $\O(\log\sigma)$. 

The suffix tree topology can be used in combination with the FM-index to support suffix links as well. 
More precisely using $\mathtt{select}$ on the \textsc{bwt}, the vector $C$ and the operations $\mathtt{lca}$, $\mathtt{leftmost\_leaf}$,
and $\mathtt{rightmost\_leaf}$ on the suffix tree topology one can deduce the interval $[i',j']$ that corresponds
to a right-maximal factor $p$, given the interval $[i,j]$ that corresponds to a right-maximal factor $cp$ (where 
$c$ is a character). First given the leaf $i$ corresponding to a suffix $s_i=cpx_i$ (where $x$ is a string), the leaf $i''$ that corresponds to the suffix $px_i$ is deduce through the formulae $i''=\mathtt{select}_c(i-C[i])$. Then the leaf $j$ corresponding to the suffix $s_j=cpx_j$ is converted to the leaf $j''$ that correspond to suffix $px_j$ by $j''=\mathtt{select}_c(i-C[i])$. Then the node $x=\mathtt{lca}(i'',j'')$ is computed and finally the interval $[i,j]$ is deduced by 
$i'=\mathtt{leftmost\_leaf}(x)$ and $j'=\mathtt{rightmost\_leaf}(x)$. 
The time to compute a suffix link is thus dominated by the time to do a $\mathtt{select}$ query which varies between $O(1)$ and $O(\log\sigma)$ depending on the implementation.

\subsubsection{Compressed suffix array}
The Compressed suffix array is a data structure that can simulate the suffix array using much less space than the original suffix array at the price of a slower access to the suffix array elements. The original suffix array occupies $n\log n$ bits of space and an access to any of its elements can be done in $O(1)$ time. 
The \textsc{CSA} instead offers the following tradeoffs: 
\begin{enumerate}
\item Space $O(n\log\sigma)$ and access time $O(\log^\epsilon n)$. 
\item Space $O(n\log\sigma\log\log n)$ and access time $O(\log\log n)$. 
\item Space $O(n\log^\epsilon n)$ time and access time $O(1)$. 
\end{enumerate}
The two first were first described in~\cite{GV05} while the latter was described in~\cite{Rao02}. 
We will restrict our interest to the first tradeoff. 

The FM-index can be considered as a special-case of a \textsc{CSA} with slower operations 
and smaller space-occupation. There exists many variants of the FM-index. The ones we use in this paper
achieve succinct space $n\log\sigma+o(n)$ with $t_{\mathtt{SA}}=O(\log n\log\log n)$ by using the wavelet tree
to represent the \textsc{bwt} or $n\log\sigma(1+o(1))$ bits of space with $t_{\mathtt{SA}}=O(\log_\sigma n\log\log n)$
by using a faster representation of the \textsc{bwt}. 

\subsubsection{Compressed suffix tree}
A compressed suffix tree~\cite{Sa07a} has three main components: 
\begin{enumerate}
\item A compressed suffix array. This component (its fastest variant) can be built in $O(n\log\log\sigma)$ time while using $O(n\log\sigma)$ bits of temporary space~\cite{HSS09}. 
%This component occupies in general at least $n\log\sigma+o(n)$ or (of even $nH_k+o(n)$ for $k=o(\log_\sigma n)$). 
\item The suffix tree topology which occupies $4n+o(n)$, but which was subsequently reduced to $2.54n+o(n)$ bits~\cite{Fi11}. 
\item The permuted lcp array (\textsc{plcp} array for short). This occupies $2n+o(n)$ bits of space. 
\end{enumerate}
The \textsc{CST} supports the same operations as the (uncompressed) suffix tree, but some important operations are supported in $O(t_{\mathtt{SA}})$ time instead of $O(1)$ in the (uncompressed) suffix tree.  

\subsubsection{Bidirectional Burrows-wheeler}
A tool that we will use is the bidirectional \textsc{bwt}~\cite{LYLLYKW09,SOG12}. 
The bidirectional \textsc{bwt} consists in two \textsc{bwt}s, one on $T[1..n-1]\$$ (which we call \textsc{bwt}) and the other on $\overline{T}[1..n-1]\$$ (which we call reverse \textsc{bwt}), where $\overline{T}$ denotes the reverse of the text $T$. In the context of the bidirectional \textsc{bwt} we will define the concept of left-maximal factors. 
The core operation of the bidirectional \textsc{bwt} consists in counting the number of occurrences of all characters smaller than $c$ in some interval $[i,j]$ of the \textsc{bwt} or the reverse \textsc{bwt}. The data structures presented in~\cite{LYLLYKW09}, \cite{SOG12} and~\cite{BCKM13} all use $O(n\log\sigma)$ bits and support the operation is respectively times $O(\sigma)$, $O(\log\sigma)$ (using the Wavelet tree) and $O(1)$. 
A factor $p$ is said to be left-maximal if and only if we have at least two distinct characters $a$ and $b$ such that the two factors $ap$ and $bp$ appear in $\$T[1..n-1]$. 

The bidirectional \textsc{bwt} has two key functionalities that will be of interest to us: the first one is the capability to efficiently enumerate all the suffix array intervals and the second one is the bidirectional navigation. 

At any given time, we maintain for every factor $p$ the suffix array interval $[i,j]$ of $p$ in the \textsc{bwt} and the suffix array interval $[i',j']$ of $\overline{p}$ in the reverse \textsc{bwt}. 

Given the factor $p$ with the suffix array interval of $p$ in the \textsc{bwt} and the suffix array interval of $\overline{p}$ in the reverse \textsc{bwt}, and a character $c$, we can recover the pair of suffix array intervals that correspond to $cp$ in the \textsc{bwt} and to $\overline{cp}$ in the reverse \textsc{bwt} in time $O(\log\sigma)$. We can also recover the suffix array intervals that correspond to $pc$ and $\overline{pc}$ in time $O(\log\sigma)$. We call the first operation $\mathtt{extendleft}$ and the second one $\mathtt{extendright}$. 

To implement the $\mathtt{extendleft}$ operation, given a character $c$, an interval $[i_1,j_1]$ corresponding to a factor $p$ in the \textsc{bwt} and an interval $[i'_1,j'_1]$ corresponding to $\overline{p}$ in the reverse \textsc{bwt}, we first use the \textsc{bwt} to get the interval $[i_2,j_2]$ that corresponds to the factor $cp$ and we let $n_c=j_2-i_2+1$ be the number of occurrences of character $c$ in the interval $\mathtt{bwt}[i_1,j_1]$. We also recover the number of occurrences of characters $b<c$ in $bwt[i_1..j_1]$. We let this count be noted by $n_b$. Then the interval $[i'_2,j'_2]$ that corresponds to $\overline{cp}$ is computed as $i'_2=i'_1+n_b$ and $j'_2=i'_2+n_c-1$. 
The operation $\mathtt{extendright}$ is symmetric to operation $\mathtt{extendleft}$ and can be implemented similarly (by changing the roles of the \textsc{bwt} and reverse \textsc{bwt}). 

We can also support two other operations called $\mathtt{contractleft}$ and $\mathtt{contractright}$. The first one operates on a right-maximal factor $cp$ (where $c$ is a character) and allows to get the pair of intervals that correspond to $p$ and $\overline{p}$ respectively. The second is symmetric and operates on a left-maximal factor $pc$ and allows to get the pair of intervals that correspond to $p$ and $\overline{p}$. Since they are symmetric, we only describe $\mathtt{contractleft}$. 

In order to support $\mathtt{contractleft}$, given the interval $[i_1,j_1]$ of $cp$ in the \textsc{bwt}, we first deduce the interval $[i_2,j_2]$
that corresponds to $p$ using the \textsc{bwt} and the suffix tree topology. We then count $n_b$, the number of occurrences of characters $b<c$ in $\mathtt{bwt}[i_2,j_2]$. From there and given the interval $[i'_1,j'_1]$ in the reverse \textsc{bwt} of the factor $\overline{cp}$, we compute $i'_2=i'_1-n_b$ and $j'_2=i'_2+(i_2-i_1)$. 
%The operation $\mathtt{contractleft}$ is symmetric to $\mathtt{contractright}$ and is implemented similarly (by using the reverse \textsc{bwt} and the topology of the suffix tree of the reverse string instead of the \textsc{bwt} and the suffix tree topology). 

\section{Construction in {\large $\boldsymbol{O(\lowercase{n}\log\sigma)}$} time \\and space}
\label{sec:small_alpha_build}

Until now it was not known how to construct the second and third components of the \textsc{CST} in better than $O(n\log^\epsilon n)$ time if the construction space is restricted to be $O(n\log\sigma)$ bits. The best time to construct the first component was $O(n\log\log\sigma)$~\cite{HSS09}. The two other components are constructed using the approach described in~\cite{HS02}. 
In this section, we show that it is indeed possible to construct them in $O(n\log\sigma)$ time. We first show how we can efficiently build the suffix tree topology based on any method that can enumerate the suffix array intervals in succinct space (for example the one in~\cite{BBO12,BGOS13} or the recent one in~\cite{BCKM13}). 
We then show how to use the bidirectional \textsc{bwt} augmented with the suffix tree topology to construct the \textsc{plcp} array. %We note that the contract capability itself needs the tree topology built in the first step. 
Our approach is different from the one taken in~\cite{HS02}, where the \textsc{plcp} array is built first (using the algorithm of~\cite{KLAAP01}), and then the suffix tree topology is induced from the \textsc{plcp} array. The main limitation of that approach came from the fact that the algorithm of~\cite{KLAAP01} needs to make expensive accesses to the suffix array and its inverse which cost $O(\log^\epsilon n)$ time per access. Moreover the construction of the tree topology needs to access the \textsc{lcp} array in natural order (not in permuted order) which again costs $O(\log^\epsilon n)$ time. 
\subsection{Building the suffix tree topology}
We first show that the suffix tree topology can be built in time $O(n\cdot t_e)$, where $t_e$ is the time needed to enumerate a suffix array interval. Typically $t_e$ will either be $O(1)$, $O(\log\sigma)$ or $O(\log\log\sigma)$. 
Our method is rather simple. Consider the balanced parenthesis representation of a suffix tree topology. Our key observation is that we can easily build a balanced parenthesis representation by enumerating the suffix array intervals. More precisely for every position in $[1..n]$, we associate two counters, one for open and the other for close parentheses implemented through two arrays of counters $C_o[1..n]$ and $C_c[1..n]$. Then given a suffix array interval $[i,j]$ we will simply increment the counters $C_o[i]$ and $C_c[j]$. Then we scan the counters $C_c$ and $C_o$ in parallel and for each $i$ from $1$ to $n$, write $C_o[i]$ opening parentheses followed by $C_c[i]$ closing parentheses. It is easy to see that the constructed sequence is that of the balanced parentheses of the suffix tree. 
It remains to show how to implement counters $C_c$ and $C_o$. A naive implementation would use $O(n\log n)$ bits of space. 
We can easily reduce the space to $O(n)$ bits of space as follows. We divide $C_o[1..n]$ into $\log\log n$ buckets (for simplicity and without loss of generality we assume that $\log n$ is a power of two) of $\log\log n$ positions each ($C_c$ and $C_o$ handled similarly, so from now on, we only describe the procedure for $C_o$). For each bucket we associate a counter of length $2\log\log n$ bits. Then we do two passes. In the first pass we increment the counter number $i/\log\log n$ every time we want to increment position number $i$. If the bucket counter reaches the value $\log^2n-1$, then we stop incrementing the counter (we call such bucket as saturated buckets). At the end of the first pass, we do the following: for every saturated bucket, we allocate a memory area of size $\log n\log \log n$, such that every position has now dedicated $\log n$ bits. For every non-saturated bucket whose counter has value $t$, we associate a memory area of size $s=\lceil \log\log n(3+2\log((t+\log\log n)/\log\log n))\rceil\leq 4(\log\log n)^2$ bits (we call the content of that memory are as bucket configuration). Note that $\log\log n(3+2\log((t+\log\log n)/\log\log n))\leq 5\log\log n+2t$ (this derives simply from the fact that $\log x\leq x$ for all $x\geq 1$). When summed up over all buckets the space becomes at most $7n$ bits~\footnote{It is likely that this constant can be improved through the use of a more efficient encoding of the counters and a tighter space analysis. Since our goal is to prove the $O(n)$ bound, we prefer to keep the encoding and the analysis as simple as possible.}. The memory area is enough to store all the counters of all the $\log\log n$ positions. For that we will use Elias-Gamma encoding~\cite{El75} that encodes an integer $x\geq 0$ using exactly $1+2\lceil\log (x+1)\rceil<3+2\log (x+1)$ bits. 
Let us denote by $x_i$ the value of the counter for position $i\in[1..\log\log n]$. Since the logarithm is a concave function we can apply the Jensen inequality to deduce that the total size of the encoding is less than: 
$$\sum_{i=1}^{\log\log n}(3+2\log (x_i+1))<\log\log n(3+2\log \frac{\sum_{i=1}^{\log\log n} (x_i+1)}{\log\log n})$$ 
and we have that:
$$\log\log n(3+2\log \frac{\sum_{i=1}^{\log\log n} (x_i+1)}{\log\log n})=\log\log n(3+2\log((t+\log\log n)/\log\log n))$$ 
We concatenate the encoding of all counters in a bucket and pad the remaining allocated bits to zero. 
This allows to have a canonical unique encoding for the counters in a bucket occupying exactly $\lceil \log\log n(3+2\log((t+\log\log n)/\log\log n))\rceil$. 

In order to efficiently support incrementation of individual counters in all bucket configurations that use the same space $s$, we will use the four-russian technique. That is for every $s\in[1..4\log\log^2 n]$, we store a table $T[2^s,\log\log n]$ where position $T[i,j]$ stores the next configuration obtained after incrementing a counter number $j$ in a bucket which had previously configuration $i$. Note that the total space used by the table is $4(\log\log n)^22^{4(\log\log)^2}\log\log n=o(n)$ bits of space and its construction can be done in time $o(n)$ as a preprocessing step. 
We concatenate the memory areas of all buckets and store a prefix-sum data structure that tells us the starting position of the memory area allocated to each bucket. This prefix-sum data structure occupies $O(n)$ bits of space and in constant time gives a pointer to the area. 

We now describe the second pass. In the second pass, we do the following. For each interval $[i,j]$ we increment the counters $C_o[i]$ and $C_c[j]$, where each counter is incremented by first looking at the prefix-sum data structures that will tell us the corresponding area. Then if the area is of size $\log n\log\log n$ we deduce that the counter is part of a saturated bucket. Otherwise, the counter is part of a non-saturated bucket. In the first case we directly increment the individual counter. In the second case, we use the lookup table to increment the counter. 

At the end we get a sequence of at most $2(2n-1)$ balanced parenthesis. We then can in $O(n)$ time build a data structure that occupies $4n+o(n)$ bits of space so as to support all operations on the topology in constant time~\cite{SNsoda10,NStalg14}. 
\begin{lemma}
\label{lemma:tree_topology}
Given a data structure able to enumerate all the suffix array intervals in $t_e$ time per interval, we can build the suffix tree topology in $O(n\cdot t_e)$ time and $O(n)$ bits of additional space. 
\end{lemma}

\subsection{Building the permuted lcp array}
The Longest common prefix array~\cite{MM93} (\textsc{lcp} array) is defined as follows: \textsc{lcp}[i]=j if and only if the longest common prefix between the suffixes of ranks $i-1$ and $i$ is equal to $j$ (the array is defined over the range $[2..n]$). The permuted \textsc{lcp} array (the \textsc{plcp} array) is defined as follows \textsc{plcp}[i]=j if and only if the rank of the suffix $T[i..n-1]$ is $r$ and the longest common prefix between that suffix and the suffix of rank $r-1$ equals $j$ (here the suffix of rank $1$ is the empty one). The \textsc{plcp} array is a permutation of the \textsc{lcp} array with the nice property that it can be encoded using only $2n$ bits~\cite{Sa02}. 
We can easily build the \textsc{plcp} array by inverting the \textsc{bwt} and using the extension and contraction capabilities. For each suffix $T[i..n-1]$ of rank $r_i$, we have to determine the largest $\ell_i$ such that $T[i..i+\ell_i-1]$ has an associated suffix array interval $[r_s,r_e]$ with $r_s<r_i$. In other words, the lcp between the suffix of rank $r_i$ and the suffixes of ranks $r_s,\ldots r_i-1$ is precisely $\ell_i$. This is evident from the fact that $T[i..i+\ell_i]$ has an associated interval $[r_s,r_e]$ with $r_s=r_i$ (the longest common prefix between the suffixes of ranks $r_s-1$ and $r_s$ is less than $\ell_i+1$). 

We use the observation that $\ell_{i-1}\leq \ell_i+1$ to devise a simple algorithm to compute $\ell_i$ starting from $i=1$ until $i=n$. 
Throughout the algorithm we will maintain two intervals: one interval in the \textsc{bwt} and the other in the reverse \textsc{bwt}. 
The algorithm works as follows: we suppose that we have $\ell_{i-1}$ with an associated pair of intervals $[r_s,r_e]$ (the suffix array interval of $T[i-1..i+\ell_{i-1}-2]$ in the \textsc{bwt}) and $[r'_s,r'_e]$ (the suffix array interval of $\overline{T[i-1..i+\ell_{i-1}-2]}$ in the reverse \textsc{bwt}) and we want to induce the two intervals that correspond to $\ell_i$ (the pair of intervals that correspond to $T[i..i+\ell_{i}-1]$ and $\overline{T[i..i+\ell_{i}-1]}$). Except when $i=1$ or when $\ell_{i-1}=0$, we first start by taking a suffix link from interval $[r_s,r_e]$ in the \textsc{bwt} and assign it to $[r_s,r_e]$. We then induce a new interval $[r'_s,r'_e]$ in the reverse \textsc{bwt}. If $i=1$ or $\ell_{i-1}=0$, we set the intervals $[r_s,r_e]$ and $[r'_s,r'_e]$  to $[1..n]$ (the interval corresponding to the empty string). 
We then do an $\mathtt{extendright}$ operation on the pair of intervals using character $T[i+\ell_{i-1}]$ (we assume that $\ell_0=0$). If that operation gives an interval $[r_s,r_e]$ with $r_s=r_i$, we stop and set $\ell_i=\ell_{i-1}-1$ (unless $\ell_{i-1}=0$ in which case we set $\ell_i=0$), otherwise we do $\mathtt{extendright}$ using character $T[i+\ell_{i-1}+1]$ and continue that way until we either reach the end of the string or reach a character $T[i+\ell_{i-1}+j]$ that gives an interval $[r_s,r_e]$ with $r_s=r_i$. In which case we set $\ell_i=\ell_{i-1}+j$. We now consider the pair of intervals $[r_s,r_s]$ and $[r'_s,r'_s]$ that correspond to $T[i..i+\ell_i-1]$ and $\overline{T[i..i+\ell_i-1]}$. If the string $T[i..i+\ell_i-1]$ is non-empty then it is evident that it is a path of a node in the suffix tree of $T$. This is because:
\begin{enumerate}
\item The suffix $T[i..n]$ starts with $T[i..i+\ell_i]$. 
\item The suffix of rank $r_i-1$ is prefixed by $T[i..i+\ell_i-1]$ (this is because the interval $[r_s,r_s]$ that corresponds to $T[i..i+\ell_i-1]$ is such that $r_s<r_i$ and thus $r_s\leq r_i-1$), but is not prefixed  
by $T[i..i+\ell_i]$ (this is because the interval $[r_s,r_s]$ that corresponds to $T[i..i+\ell_i-1]$ is such that $r_s=r_i$ and thus $r_i-1<r_s$). Thus the suffix of rank $r_i-1$ starts with $T[i..i+\ell_i-1]$, but is followed by a character different from $T[i+\ell_i]$. 
\end{enumerate}
Thus we have that $T[i..i+\ell_i-1]$ is right maximal as the factor $T[i..i+\ell_i-1]$ has at least two occurrences in which it is followed by two distinct characters. We can use $\mathtt{contractleft}$ to induce the pair of intervals that correspond 
to the factors $T[i+1..i+\ell_i-1]$ and $\overline{T[i+1..i+\ell_i-1]}$. We can then do a sequence of $\mathtt{extendright}$ operations to compute $\ell_{i+1}$ 
exactly in the same way as we computed $\ell_i$. 

%We first determine the two intervals $[r_s,r_e]$ (in bwt) and $[r'_s,r'_e]$ (in reverse bwt) that correspond to character $T[1]$ and we suppose that we know the rank $r_1$ of the suffix $T[1..n]$. Then, we check whether $r_1=r_s$, in which case $\ell_1=0$. This is evident from the fact that 

%The strategy works as follows: suppose we have determine $\ell_i$, then we determine $\ell_{i-1}$ by first checking whether $T[i-1..i-1+\ell_i]$ has an associated interval $[r_s,r_e]$ with $r_s<r_{i-1}$. If this is the case, we deduce that $\ell{i-1}=\ell_i+1$. Otherwise, we successively check whether $T[i-1..i-1+\ell_i-j]$ has an associated interval $[r_s,r_e]$ with $r_s<r_{i-1}$ for increasing values of $j\leq 1$ until we get $r_s=r_{i-1}$ in which case we get that $\ell_{i-1}=\ell_i-j+1$. Notice that the total time is the time taken to do $n$ extensions and at most $n$ contractions, thus taking time $O(n\log\sigma)$ or $O(n)$ depending on the used data structure. 

\begin{lemma}
\label{lemma:plcp}
Given a bidirectional text index built on a text of length $n$ and that that allows $\mathtt{extendleft}$ and $\mathtt{contractright}$ queries in time $t$, we can build the \textsc{plcp} array in $O(t\cdot n)$ time and $O(n)$ bits of 
additional space. 
\end{lemma}
We can now combine Lemma~\ref{lemma:plcp} and Lemma~\ref{lemma:tree_topology} with the use of a wavelet tree as a representation of the bidirectional \textsc{bwt} which allows operations $\mathtt{extendleft}$ and $\mathtt{contractright}$ and the enumeration of intervals in time $O(\log\sigma)$ resulting in the following theorem. 
\begin{theorem}
Given a string of length $n$ over an alphabet of size $\sigma$, we can build the three main components of the compressed suffix tree in $O(n\log\sigma)$ time and bits of space. 
\end{theorem}

%\subsection{applications}
%One immediate application is that the matching statistics between two strings $s$ and $t$ can be obtained in $O((|s|+|t|)\log\sigma)$ time and space. Previous result either needed $\omega(|s|+|t|)\log\sigma)$ bits of space or time. 

\section{Construction in linear time\\ and {\large $\boldsymbol{O(\lowercase{n}\log\sigma)}$} space}
\label{sec:big_alpha_build}

We can show that we can construct both the \textsc{CST} and the \textsc{CSA} in $O(n)$ (randomized) time while still using $O(n\log\sigma)$ bits of additional space. 
%We first start by showing some basic building blocks. 

The following lemma is shown in appendix~\ref{sec:build_range_color_rep}. 
\begin{lemma}
\label{lemma:build_range_color_report1}
Given a sequence $S[1..n]$ of colors from $[1..\sigma]$ represented 
using a data structure that allows randomly accessing any element 
of the sequence in time $t_{\mathtt{access}}$, we can 
in randomized $O(n)$ time build a data structure that occupies 
$O(n\log\log\sigma)$ bits so that 
given any range $[i,j]$, we can report all the $\mathtt{occ}$ distinct
colors that occur in $S[i..j]$ in time $O(\mathtt{occ}(1+t_{\mathtt{access}}))$, 
In addition, for every color occurring in $S[i..j]$, the algorithm reports 
the frequency of the color in $S[i..j]$ and in $S[1..i-1]$
The reporting algorithm uses $O(\sigma)$ bits of working space. 
\end{lemma}
The data structure which was described in~\cite{BNV13} combines the use of a 
range color reporting structure, that is used to report the rightmost 
and leftmost occurrences of every distinct color in a range $[i,j]$ with $\sigma$
different $\textsc{mmphf}$s that allow to compute the ranks 
of the leftmost and rightmost occurrences of the reported characters
in the range.
The frequency in $S[1..i-1]$ is obtained by subtracting one 
from the rank of the leftmost occurrence and the frequency in $S[i..j]$ is one 
plus the difference between the two ranks. 

We will present a different result: 
\begin{lemma}
\label{lemma:build_prank}
Given as input sequence $S[1..n]$ of colors from $[1..\sigma]$, 
we can in deterministic $O(n)$ time 
build a data structure that occupies $n\log\sigma+4n+o(n)$
bits of space and that supports $\mathtt{access}$ and 
$\mathtt{prank}$ operations in constant time. That is, 
given a position $i$ the data structure returns the color 
$c=S[i]$ the count $\mathtt{rank}_c(i)$ in constant time. 
\end{lemma}
The result is obtained by borrowing some ideas from~\cite{GMR06}. 
We divide the sequence $S[1..n]$ into chunks of length $\sigma$ 
characters each. That is, into blocks $S[1..\sigma]$, 
$S[\sigma+1,2\sigma]$ and so on. Then for every 
character $c$ store a bitmap $V_c$ that tells how many occurrences 
of the character $c$ are in each block (store the sequence 
$10^{\mathtt{freq}_{(c,1)}}10^{\mathtt{freq}_{(c,2)}},\ldots,10^{\mathtt{freq}_{(c,n/\sigma)}}$, 
where $\mathtt{freq}_{(c,i)}$ gives the frequency of character $c$
in block $i$). The total length 
of the bitmaps for all characters is $2n+o(n)$
Then for every block $i$ (denote it by $S_i$) store:
\begin{enumerate}
\item a bitvector $B_i$ of length $2\sigma$ that contains 
$10^{\mathtt{freq}_{(1,i)}}, 10^{\mathtt{freq}_{(2,i)}},\ldots,10^{\mathtt{freq}_{(\sigma,i)}}$, 
where $\mathtt{freq}_{(c,i)}$ gives the frequency of character $c$
in block $i$. 
\item Store a sequence $S'_i[1..\sigma]$ in which 
$S'_i[j]$ stores the partial rank of $S_i[j]$ in $S_i$
(the number of occurrences of character 
$c=S_i[j]$ inside $S_i[1..j]$) added to the number 
of occurrences of characters $b<c$ in $S_i$. 
\end{enumerate}
Now a partial rank query for position $j$ in $S$ is easy 
to resolve: first retrieve $x=S'[i]$, then the character $c=s[i]$
is given by $\mathtt{select}_0(B_i,x)-x$.
Finally the partial rank of $c$ in $S_i$  is given by $\mathtt{select}_0(B_i,x)-\mathtt{select}_1(B_i,c)$. 
Finally the partial rank in $S$ is given by adding this partial 
rank to the count of $c$ in the blocks $[1..i-1]$ of $S$ which is retrieved 
using select on bitvector $V_c$. Overall the space is $n\log sigma+4n+o(n)$ bits. 

The only source of randomization construction time in Lemma~\ref{lemma:build_range_color_report1}
is in the construction of $\textsc{mmphf}$s used to support the partial rank 
queries. If we instead use partial rank data structure from Lemma~\ref{lemma:build_prank} 
into Lemma~\ref{lemma:build_range_color_report1}, we immediately get the following 
corollary: 
\begin{lemma}
\label{lemma:build_range_color_report2}
Given as input sequence $S[1..n]$ of colors from $[1..\sigma]$, 
we can in deterministic linear time 
build a data structure that occupies $n\log\sigma+8n+o(n)$
bits of space and that supports $\mathtt{access}$ and 
$\mathtt{prank}$ operations in constant time
and in addition, allows, given any range $[i,j]$ to 
report all the $\mathtt{occ}$ distinct
colors that occur in $S[i..j]$ in time $O(\mathtt{occ})$. 
In addition, for every color occurring in $S[i..j]$, the algorithm reports 
the frequency of the color in $S[i..j]$ and in $S[1..i-1]$
The reporting algorithm uses $O(\sigma)$ bits of working space. 
\end{lemma}

\subsection{Interval enumeration in linear time and compact space}
\label{subsec:linear_time_bwt_interval_enum}
We now show a method to enumerate the suffix array intervals in constant 
time per interval. For that we will use the method described in
~\cite{BBO12,BGOS13}. This method enumerates every interval in time 
$O(\log\sigma)$. The algorithm uses the \textsc{bwt} represented 
using a Wavelet tree in addition to an auxiliary bit-vector and a queue 
which occupy $O(n)$ bits of additional space. 

The bottleneck in the algorithm is the following operation. Given a
suffix array interval $[i,j]$ find all $d$ distinct characters that appear in $\mathtt{bwt}[i,j]$
and for each such character $c$ compute a Weiner link from $[i,j]$. 
The latter operation amounts to computing the number of occurrences of $c$ in $\mathtt{bwt}[1..i-1]$ 
and in $\mathtt{bwt}[i..j]$. Using a Wavelet tree to represent the \textsc{bwt}, we can compute 
those numbers in $O(d\log\sigma)$ time~\cite{BGOS13}. 
We instead use Lemma~\ref{lemma:build_range_color_report1} 
or Lemma~\ref{lemma:build_range_color_report2} to support the operation in constant time
per reported color. 
We can further show a modified version that uses a stack instead of a queue. 
The stack occupies $O(\sigma^2\log^2n)=O(n^{2/3}\log n)=o(n)$ bits.

The general idea of the algorithm is to enumerate all the suffix array intervals that correspond
to suffix tree nodes through the use of explicit Weiner links starting 
from the root of the tree. For every node, we maintain the sub-intervals that correspond 
to its children. Then, there will be an explicit Weiner link from the node labeled 
with character $c$, if and only 
if there exist Weiner links from $a\geq 2$ children of the node
labeled with the same character $c$. Moreover this target node will have 
exactly $a$ children. 
Otherwise the existence of only one Weiner link labeled with character $c$ from one child, 
indicates an implicit Weiner link from the parent node. Since both the total number of internals nodes 
with their children and the total number of implicit Weiner links are linear, we deduce that the 
total number of Weiner links computed by the algorithm is also linear. 

We now show the details of the algorithm. 

%Here we show that we can do the same in $O(d)$ time only by using the method of~\cite{BNV13}
%which can given an interval $[i,j]$ enumerate all the $d$ distinct colors that occur in the interval 
%and count their frequencies in time $O(d)$. The method uses the following idea which was first proposed
%in~\cite{Sa07b}. We use two \textsc{rmq}s on top of the sequence of colors. The first one 
%allows to enumerate all the leftmost occurrences of all distinct colors. The second allows to enumerate
%all the rightmost occurrences of the colors. We then use a monotone minimal perfect hash function 
%to compute the rank of the leftest occurrences and the rank of the rightest occurrence of each color. 
%The frequency of the color is obtained by subtracting the first rank from the second and incrementing 
%the result by $1$. 

We use three vectors. The first vector $V[1..\sigma][1..\sigma]$
stores pairs $(c,[i_c,j_c])$ where $c$ is a character and $[i_c,j_c]$
is an interval of integers. The second vector $Y[1..\sigma]$ stores integer counters
initially set to zero. We finally have a vector of characters $W[1..\sigma]$ and 
an integer counter $N_W$ associated with it and initially set to zero. 

We start by enumerating the $d$ children of the root, which 
can be done directly using the array $C$. We let $\alpha_1,\ldots,\alpha_d$
be the child labels in sorted order and their associated 
intervals $[i_1,j_1],[i_2,j_2],\ldots,[i_d,j_d]$. For every $x=1,2,\ldots,d$, 
we enumerate all the characters $c_1,c_2,\ldots c_k$ that occur in $\mathtt{bwt}[i_x,j_x]$ and 
for each character $c_y$ compute the associated interval $[i_y,j_y]$ which is the 
interval of target node of the Weiner link labeled with $c_y$ 
and starting from the node associated to the interval $[i_x,j_x]$. 
We then increment the counter $Y[c_y]$
and store the pair formed by $\alpha_x$ and the interval $[i_y,j_y]$ in $V[c_y][Y[c_y]]$. 
If the counter $Y[c_y]$ has now value $1$ we append $\alpha_x$ 
at the end of $W$, by
incrementing $N_W$ and setting $W[N_W]=c_y$. 

We now scan the vector $W$ and for each $c\in W[1..N_W]$ with $Y[c]>1$,
consider all pairs in $V[c][1..Y[c]]$. Then $c$ will be the path 
labeling a node in the suffix tree and each pair $(\alpha,[i_\alpha,j_\alpha])$
in $V[c][1..Y[c]]$ represents the label of a child $\alpha$ 
and its interval in the \textsc{bwt} $[i_\alpha,j_\alpha]$. The reason why $c$ 
is the label of a path is easy to see. Since there exist at least two distinct characters $\alpha_x$
and $\alpha_y$ in  $V[c][1..Y[c]]$ means that we have at least two children 
of the root labeled by $\alpha_x$ and $\alpha_y$ which have Weiner links labeled by $c$. 
Thus the paths of the two targets of the two Weiner links are two nodes whose paths are prefixed 
respectively by $c\alpha_x$ and $c\alpha_y$. Thus we deduce that $c$ is right-maximal. 

Note that the characters in the pairs $V[c][1..Y[c]]$ are in sorted order. 
That is $V[c][1..z]$ with $z=Y[c]$ contains the pairs 
$(\alpha_1,[i_{\alpha_1},j_{\alpha_1}])\ldots (\alpha_z,[i_{\alpha_z},j_{\alpha_z}])$, 
where $\alpha_1<\alpha_2<\ldots <\alpha_z$. Then the interval corresponding to 
path $c$ will be $[i_{\alpha_1},j_{\alpha_z}]$. For each $c$ we will 
compute the interval size $u_c=j_{\alpha_z}-i_{\alpha_1}+1$. 

So at the first step we will have deduced all internal nodes labeled with paths 
of length $1$. For each such node, we will have its interval, the labels 
of all its children along with their sub-intervals. 
We will use the stack in order to recursively enumerate 
all the internal nodes labeled by paths of lengths greater than $1$. 
For every node with a path of length $1$, labeled by path $c$ we will 
push the array of pairs $V[c][1..Y[c]]$ (we prepend the array by its length $Y[c]$)
and reset $Y[c]$ to zero. 
However we will make sure to push the array of the node with the largest interval size $u_c$ first. 
This will ensure that the stack will contain at most $\sigma\log n$ arrays at any point in time~\footnote
{This trick is used in~\cite{BCKM13} and was already used as early as the quicksort 
algorithm~\cite{hoare1962quicksort}.}. 
We then pop the array on top of the stack and then enumerate all the nodes with path of length $2$
reachable using an explicit Weiner link from the node corresponding to that array
using exactly the same method that was used to induce the nodes reachable by explicit Weiner links from 
the root. We push the corresponding arrays on the stack, pop the one on the top
and continue that way. 

At the end, the algorithm will enumerate all the intervals that correspond to internal 
nodes along with the child labels and their corresponding sub-intervals. 

The space usage of the stack will be bounded by $O(\sigma^2\log^2n)$ bits of space, since the 
depth of the stack is $\log n$ and for every internal node, we push at most $\sigma$ arrays 
corresponding to explicit Weiner links from that node and each array is of size $O(\sigma\log n)$ bits. 
Given that $\sigma\leq n^{1/3}$, we deduce that the total space used by the stack will be $o(n)$ bits
of space. 

\begin{lemma}
\label{lemma:enum_intervals}
Given a text $T$ of length $n$ over an alphabet of size $\sigma$ and for which we have already built the \textsc{bwt}  
and supposing that we have built a data structure on top of the \textsc{bwt} that can enumerate the distinct characters 
in any interval along with the rank of their leftest and rightest occurrences, in time $O(t_e)$ per character. 
We can then in time $O(n\cdot t_e)$ time,  using $o(n)$ bits of additional space, enumerate all the suffix array intervals that correspond to the internal nodes  in the suffix tree of $T$ and for every node, enumerate the labels of its children along with their corresponding sub-intervals in sorted lexicographic order of their labels. 
\end{lemma}

By combining Lemma~\ref{lemma:build_range_color_report2} with Lemma~\ref{lemma:enum_intervals} and plugging the result into Lemma~\ref{lemma:tree_topology}, we immediately get the following lemma. 

\begin{lemma}
\label{lemma:tree_topology2}
Given a text $T$ of length $n$ over an alphabet of size $\sigma$ and for which we have already built the \textsc{bwt}, we can build the suffix tree topology in deterministic $O(n)$ time using $O(n)$ bits of additional space. 
\end{lemma}

\subsection{Linear time construction of the Burrows-Wheeler transform}
\label{subsec:linear_time_bwt_build}

We will now show that the \textsc{bwt} can be built in randomized
linear time. We use the recursive approach of~\cite{HSS09}. Given a text $T[1..n-1]\$$,
we let $B=\lfloor\log_\sigma n/3\rfloor$. We let $T'[1..B\cdot \lceil n/B\rceil]$ be
the text obtained by appending $T[1..n-1]\$$ with $B\cdot \lceil n/B\rceil-n\leq B-1$ 
occurrences of character $\$$ (this makes sure the length of $T'$ is multiple of $B$). 
We let $n'$ be the length of $T'$.

We form a new text $T_B[1..n'/B]$, by first grouping every 
block of $B$ consecutive characters of $T'$ into a single one.

We then build the \textsc{bwt} of $T_B$ using any linear time
algorithm for building suffix arrays~\cite{KA03,KSPP05,KSB06}
and replacing each suffix pointer 
by the character that precedes the suffix (in the cyclic rotation of $T_B$). 
This will use $O(n/B)=O(n/\log_\sigma n)$ time and 
$O((n/B)\log (n/B))=O(n\log\sigma)$ bits of space.

We now consider the string $T^r_B$ obtained by rotating 
$T'$ to the left by $B/2$ characters and then grouping every $B$
consecutive characters of the obtained text into a single character 
(intuitively we are considering $n'/B$ suffixes of $T'$ that start at positions 
$B/2+1,3B/2+1,\ldots,n'/B-B/2+1$). We can induce the \textsc{bwt} of $T^r_B$ 
from the \textsc{bwt} of $T_B$ as shown in~\cite{OS09} in time $O(n'/B)$.

We then consider $T_{B/2}$ which is constructed by grouping every block 
of $B/2$ consecutive characters of $T'$ into a single character. 
In order build the \textsc{bwt} of $T_{B/2}$ we will merge 
the \textsc{bwt} of $T_B$ and $T^r_B$. We then replace every character in the merged \textsc{bwt} 
with its second (right) half. The obtained result is clearly
the \textsc{bwt} of $T_{B/2}$. 
%We then 
%build the \textsc{bwt} of $T_{B/2}$ in two phases (as shown in ~\cite{HSS09}). 
%In the first phase we construct the subsequence of the \textsc{bwt} 
%that contains only characters preceding cyclic rotations of $T_{B/2}$ that 
%start at odd positions (equivalently suffixes 
%$T'[1..n],T'[B+1..n]$,$T'[2B+1..n],\ldots$) and the subsequence of the 
%\textsc{bwt} that contains characters preceding cyclic rotations that start at even positions
%(equivalently suffixes $T'[B/2+1..n],T'[3B/2+1..n],\ldots$). The 
%first subsequence is easy to get from the \textsc{bwt} of $T_{B}$ since. 
%
%We can induce the second 
%subsequence in the following way. We traverse the bwt of $T_{B}$ 

We now show how to merge two \textsc{bwt}s 
into one by doing a traversal of the nodes in the virtual suffix tree 
that would contain all the suffixes in the two sets of suffixes (that correspond
to the two \textsc{bwt}s) using a variant of the method described 
in section~\ref{subsec:linear_time_bwt_interval_enum}. 
We call a node of the tree pure if all the leaves in its subtree 
come from the same set of suffixes. We call it hybrid otherwise. 
For every leaf $x$ we consider its ancestor node $y$ ($y=x$ is possible)
such that $y$ is pure and the parent of $y$ is hybrid. 
It is clear that such a node $y$ always exists, since $x$ is itself pure and 
the root is hybrid. It is also clear that such a node is unique since none of 
its ancestors is pure and none of its descendants which are ancestors of 
$x$ have a hybrid parent. 

The traversal only traverses hybrid nodes and enumerates the children 
of each of them. We note that taking a suffix link from a hybrid node
leads to a hybrid node as well. Thus all hybrid nodes are connected through 
suffix links (up to the root) 
and thus they can be enumerated by taking reverse suffix links
(explicit Weiner links) from the root. 

The traversal is done by simultaneously enumerating suffix array 
intervals for the two \textsc{bwt}s using essentially the same algorithm 
described in section~\ref{subsec:linear_time_bwt_interval_enum}. 
Such a synchronized traversal has already been described in~\cite{BCKM13}, 
but for the bidirectional \textsc{bwt}. 
During the traversal, we will use two stacks instead of one. The 
two stacks will always be synchronized in the sense that the array 
on top of the two stacks will always correspond to the node 
with the same path. 
We will also use two instances $V_1$ and $V_2$ of vectors  
$V[1..\sigma][1..\sigma]$ and two instances $Y_1$ and $Y_2$ of 
$Y[1..\sigma]$, one for each \textsc{bwt}. However we will still 
use one single instance of the vector $W$ and of the 
counter $N_W$ which will indicate us the characters $c$ for which
$t=Y_1[c]\neq 0$ or $t'=Y_2[c]\neq 0$. 
One major difference with the previous variant of the 
traversal is the choice on whether to push an array $V[c]$ on the stack or not. 
Assuming that the pairs $(\alpha_1,[i_{\alpha_1},j_{\alpha_1}]),\ldots..\alpha_t,[i_{\alpha_t},j_{\alpha_t}])$
appear in the first array $V_1[c]$ and 
$(\beta_1,[i_{\beta_1},j_{\beta_1}]),\ldots..\beta_t,[i_{\beta_{t'}},j_{\beta_{t'}}])$ 
appear in the second array $V_2[c]$. 
We require that:
\begin{enumerate} 
\item If $t=t'=1$, then $\alpha_1\neq \beta_1$. This ensure that the path 
is that of a right-maximal and thus that the two arrays indicate 
a node in the suffix tree. 
\item $t\geq 1$ and $t'\geq 1$. That indicates that the node is hybrid. 
\end{enumerate}
If the two conditions are fulfilled, then we push both arrays on
the two respective stacks. 

Given a node whose path is $p$ and which has 
one interval in first \textsc{bwt} noted $[i,j]$ and another interval in second 
\textsc{bwt} noted $[i',j']$, we will have the list of characters 
$\alpha_1<\alpha_2<\ldots <\alpha_{t}$ such that for every $x\in[1..t]$, 
there exists a sub-interval $[i_{\alpha_x},j_{\alpha_x}]$ of $[i,j]$ 
corresponding to path $p\alpha_x$.  
We will also have the list of characters 
$\beta_1<\beta_2<\ldots <\beta_{t'}$ such that for every $x\in[1..t']$, 
there exists a sub-interval $[i'_{\beta_x},j'_{\beta_x}]$ of $[i',j']$ 
corresponding to path $p\beta_x$.  

We then simultaneously traverse the lists of characters $\alpha_x$ 
and characters $\beta_x$ in increasing order. Every time we encounter 
a $\alpha_x$ that does not appear in characters $\beta_x$, we will 
deduce that the sub-interval $[i_{\alpha_x},j_{\alpha_x}]$ corresponds 
to a pure node that contains only leaves that correspond to suffixes 
in the first set. 
We then consider the largest $\beta_y<\alpha_x$ (if it exists) with its corresponding 
sub-interval $[i'_{\beta_y},j'_{\beta_y}]$. We then deduce that the suffixes 
in the sub-interval $[i_{\alpha_x},j_{\alpha_x}]$ have ranks
$[j'_{\beta_y}+i_{\alpha_x},j'_{\beta_y}+j_{\alpha_x}]$ among the union
of the two sets of suffixes. We thus place the characters that occur 
in the positions $[i_{\alpha_x},j_{\alpha_x}]$ of the first \textsc{bwt} in 
the positions $[j'_{\beta_y}+i_{\alpha_x},j'_{\beta_y}+j_{\alpha_x}]$ 
of the combined \textsc{bwt}. 

In case $\beta_y$ does not exist, then the suffixes 
in the sub-interval $[i_{\alpha_x},j_{\alpha_x}]$ have ranks
$[i'-1+i_{\alpha_x},i'-1+j_{\alpha_x}]$ among the union
of the two sets of suffixes. We thus place the characters that occur 
in positions $[i_{\alpha_x},j_{\alpha_x}]$ of the first \textsc{bwt} in 
positions $[i'-1+i_{\alpha_x},i'-1+j_{\alpha_x}]$ of the combined \textsc{bwt}. 
We handle the case where $\beta_x$ does not appear in the list of $\beta_y$
in a symmetric way. 

Once we have obtained the \textsc{bwt} of $T'_{B/2}$, 
we can reuse the same procedure to deduce the \textsc{bwt}
 of $T'_{B/4}$. We continue that way until we get 
the \textsc{bwt} of $T'_{1}=T'$. At the end we can deduce the 
\textsc{bwt} of $T$ by removing all but one occurrence of 
character $\$$~\footnote{The characters are all clustered together.}. 
\begin{lemma}
~\label{lemma:linear_bwt}
Given a string of length $n$ over an alphabet of size $\sigma$, we can build the \textsc{bwt}
 of the string in deterministic $O(n)$ time and $O(n\log\sigma)$ bits of space. 
\end{lemma}
\subsection{Completing the constructions}
The following theorem is a direct consequence of Lemma~\ref{lemma:linear_bwt} (see~\cite{HSS09}).
\begin{theorem}
\label{theo:build_bwt}
Given a string of length $n$ over an alphabet of size $\sigma$, we can build the compressed suffix array and the FM-index in (randomized) $O(n)$ time and $O(n\log\sigma)$ bits of space. 
\end{theorem}
In appendix~\ref{sec:build_ext_fm_indices}, 
we show that the recently proposed variants of the FM-index, efficient for large alphabets can also be built in randomized linear time. In particular we will show that the index of~\cite{BNtalg14} which uses $n\log\sigma(1+o(1))$ bits of space supports Weiner links in constant time, can be built in randomized linear time. The main idea of that index was to avoid the use
of the slow $\mathtt{rank}$ operation and instead simulate Weiner links using a combination $\mathtt{select}$ operation with $\textsc{mmph}$ and operations on the suffix tree topology. 
We will also show that the bidirectional FM-index index proposed in~\cite{BCKM13} can also be built in randomized linear time. 

With the help of theses indices, we will also prove in appendix~\ref{sec:build_cst} 
that the \textsc{CST} can also be built in randomized linear time, resulting in the following theorem :

\begin{theorem}
\label{theo:cst_large_alphabet}
Given a string of length $n$ over an alphabet of size $\sigma$, we can build the compressed suffix tree in (randomized) $O(n)$ time and $O(n\log\sigma)$ bits of space. The resulting compressed suffix tree occupies $O(n\log\sigma)$ bits of 
space and supports all operations in constant time, except for the string depth, the child and the string level ancestor queries which are supported in $O(\log^\epsilon n)$ time. 
\end{theorem}
The first two components of the \textsc{CST} can be built in randomized $O(n)$ time by lemmata~\ref{theo:build_bwt} and~\ref{lemma:tree_topology2}. 
The third component can be built by first constructing the bidirectional FM-index and combining it with Lemma~\ref{lemma:plcp}. 
Finally the support for child and string level ancestor queries is obtained by augmenting the \textsc{CST} with some auxiliary data structures. 

In appendix~\ref{sec:det_building}, we show that some text indexes can be built in deterministic $O(n)$ time if we allow a slight slowdown in query time. 

\section{Applications}
\label{sec:applications}
We can now use the structures proposed in~\cite{BCKM13}  
to solve many sequence analysis problems in randomized $O(n)$ time
and $O(n\log\sigma)$ bits of space which is optimal in the size of the input 
strings up to a constant factor. 
Among those problems, we mention the maximal repeats in a string, the maximal unique 
and maximal exact matches between two strings, computing the number of distinct k-mers in 
a string and many others. 
The bottleneck in those algorithms is building a bidirectional 
BWT index which previously took $O(n\log^\epsilon n)$ time when the space is limited  
to $O(n\log\sigma)$ which can now be done in linear randomized time. 
%This can now can be done in time $O(n)$ using theorem~\ref{theo:full_double_BWT}. 

We note that all the problems can directly be solved in deterministic $O(n)$ time using the traversal 
technique described in sections~\ref{subsec:linear_time_bwt_interval_enum} and~\ref{subsec:linear_time_bwt_build} 
in combination with Theorem~\ref{theo:build_bwt} instead of building the bidirectional index. 
This is because all these problems rely on the enumeration of the suffix array intervals 
that correspond to all suffix tree nodes, along with their child edges and Weiner links, 
which can now be done efficiently using the new technique. 
Using previous enumeration techniques (for example the one in~\cite{SOG12})
the best time within $O(n\log\sigma)$ bits would be $O(n\log\sigma)$. 

We believe that other kinds of sequence analysis problems 
that do not rely on the enumeration of suffix array intervals 
can also be solved in randomized linear time. In particular, those 
for which there exists solutions that rely only on Weiner links and 
on operations on the suffix tree topology.

\section*{Acknowledgements}
The author wishes to thank Veli M\"{a}kinen, Alexandru Tomescu and Travis Gagie for their valuable comments 
and encouragements and Fabio Cunial for the fruitful discussions on the subject. 
He also wishes to thank Enno Ohlebusch for his useful comments and remarks. 
\bibliographystyle{plain}
\bibliography{cst_build}

\newpage
\appendix

\section{additional tree topology operations}
In addition the tree topology supports the following $3$ operations: 
\begin{enumerate}
\item Given a node $x$ returns its parent $y$.  
\item Given a node $x$ and an index $i$, return $y$ the child number 
$i$ of $x$. The operation $\mathtt{first\_child}$ is a special case 
of the $\mathtt{child}$ operation that returns the first child of a given node $x$. 
\item Given two nodes $x$ and $y$, return $z$, the lowest common ancestor
(\textsc{lca}) of $x$ and $y$. 
\item Given a node $x$ return the indexes $i+1$ ($j+1$) 
of the leftmost leaf $y$ (rightmost leaf $z$) 
in the subtree of $x$, where $i$ ($j$)
is the number of leaves of the tree that lie respectively 
on the left of $y$ and $z$. 

\item Given a node $x$, returns its depth (distance from the root) or 
its height (the distance to its deepest descendant). 
\item Given a node $x$ and a depth $i$, the level ancestor 
query returns the ancestor of $x$ at depth $i$. 
\item Given a node $x$ return its $\mathtt{next\_sibling}$ if it exists. 
The next sibling of the node $x$ is the node $y$ such that if $x$ is the 
child number $i$ of its parent, then $y$ will be the child number $i+1$. 
\end{enumerate}

\section{Building components}
\subsection{Building the access/rank/select structures}
\label{sec:build_rank}
We show the following easy lemma which shows that we can efficiently build the structures that support $\mathtt{access}$, $\mathtt{rank}$ and $\mathtt{select}$ over large alphabet. 
%The following easy lemma is proved in appendix~\ref{sec:build_rank}
\begin{lemma}
\label{lemma:rank_select_access}
Given a sequence of length $n$ over an alphabet of size $\sigma$, we can in randomized $O(n)$ time build a data structure that occupies $n\log\sigma(1+o(1))$ bits of space and that supports $\mathtt{rank}$ in $O(\log\log\sigma)$ time and one of $\mathtt{access}$ and $\mathtt{select}$ in constant time and the other in $O(\log\log\sigma)$ time. 
\end{lemma}

The basic idea is to cut the sequence $A$ of length $n$ into $N=\lceil n/\sigma\rceil$ blocks of size $\sigma$ (except possibly for the last block which might be smaller). That is block number $i<N$ covers the subarray $A[\sigma(i-1)+1,\sigma i]$ and the last block $i=N$ covers the subarray $A[\sigma(i-1)+1,n]$. 
We then for each character $c\in[1..\sigma]$ build an bitvector $B_c$ of size $f_c+N$ bits, where $f_c$ is the number of occurrences of character $c$ in the sequence $A$. The bitvector is built as follows, we scan the blocks from left to right where for each block $i=1,2\ldots N$, we write a one followed by $f_{c,i}$ zeros, where $f_{c,i}$ is the number of occurrences of character $c$ in $A[\sigma(i-1)+1,\sigma i]$ (or $A[\sigma(i-1)+1,n]$ if $i=N$). 

The building of the two bitvectors proceeds in two phases. In the first phase we just compute the frequencies $f_c$ for all characters $c\in[1..\sigma]$. This can be done by scanning the sequence $A$ and incrementing the counter $f_c$ each time we encounter the character $c$. As we know the frequencies, we can immediately determine the size of each bitvector $B_c$ and allocate a global bitarray that will contain the concatenation of all the bitvectors. Storing the counters $f_c$ takes $O(\sigma\log n)$ bits. 
In the second phase we scan again the bitvector $f_c$ and for block $i$ append a one at the end of every vector $B_c$ and then for each character $c=A[i\sigma(i-1)+j]$ of the block append a zero to bitvector $B_c$.

The total space occupied by all bitvectors $B_c$ will be $O(n)$. That is, for each $c$, $B_c$ contains exactly $N$ ones 
which translates into $\sigma\lceil n/\sigma\rceil\leq n+\sigma-1$ bits. The number of zeros in all bitvectors $B_c$ will be the sum of $f_c$ for all $c\in[1..\sigma]$ which is exactly $n$. 
Thus the total space occupied by all bitvectors $B_c$ will be at most $2n+\sigma-1$ bits. 
The total space used by $f_c$ and by the pointers to each $B_c$ will be $O(\sigma\log n)=O( n^{1/3}\log n)=o(n)$. 
The total time used to build the two bitvectors will be $O(n)$ and the building space will be $O(n)$  in addition to the sequence itself. 

The vectors $B_c$ will allow us to reduce \textsc{rank} and \textsc{select} queries on $A$ to respectively \textsc{rank} and \textsc{select} queries on a single block of $A$ of size at most $\sigma$ (see for example~\cite{GMR06} for details). 
The data structures that support \textsc{rank} and \textsc{select} queries on a block can be built in time $O(\sigma)$
and $O(\sigma\log\sigma)\leq O(n\log\sigma)$ bits of space. For supporting $\mathtt{rank}$ queries for a given character $c$, 
a predecessor data structure, a y-fast trie~\cite{willard1983log} is built on sampled positions of occurrences of character $c$. More precisely one samples, every $\log\sigma$ positions of occurrences of character $c$ are stored in the predecessor data structure. The y-fast trie occupies $O(m\log u)$ bits and answers queries in time $O(\log\log u)$, where $m$ is the number 
of elements and $u$ is the size of the range (universe). Since the universe of positions is $\sigma=u$ and the total number of elements in all structures is $O(n)$, we deduce that the total space used by all y-fast tries will be $O((\sigma\log u)/\log\log\sigma)$. 

Then a succinct SB-tree ~\cite{grossi2009more} is built on every block of $\log\sigma-1$ non-sampled positions. Each SB-tree will use $O(\log\sigma\log\log\sigma)$ bits and allows predecessor search in constant time on the stored positions plus the time needed to do a constant number of select queries.  
Then, the query $\mathtt{rank}_c(i)$ translates into the predecessor query, which determines an interval of $\log\sigma$ occurrence positions and the predecessor search is completed by doing a predecessor query on the block of positions using the 
SB-tree. The total time will be $O(\log\log\sigma)$ plus the time to do a constant number of select queries. 
The time to build the y-fast trie data structure is $O(f_{c,i})$. The time to build every SB-tree is also $O(\log\sigma)$ which sums up to $O(f_{c,i})$ for all succinct SB-trees of a given character $c$. Actually, a succinct SB-tree is a tree constant depth and is made of nodes built on $O((\log\sigma)^{1/2})$ elements. Every SB-node uses a blind compacted trie succinctly encoded in $O((\log\sigma)^{1/2}\log\log\sigma)$ and the construction of the trie is done in linear $O((\log\sigma)^{1/2})$ time. The SB-tree needs to use a global lookup up table of size $o(\sigma)$ bits that is shared between all the the SB-trees. 
The table can be built in time $o(\sigma)$. 
Thus the total construction time for one SB-tree will be $O(\log\sigma)$ and for all SB-trees for a given character, it will be $O(f_{c,i})$.

The access query naturally translates into an local access query on a block. 
That is the query $\mathtt{access}(i)$ translates into query $\mathtt{access}(i-\sigma(\lceil i/\sigma\rceil-1))$ on block number $\lceil i/\sigma\rceil$ of $A$. 

It remains to show how \textsc{select} and \textsc{access} queries are supported on a block $b_i[1..\sigma]=A[(i-1)\sigma,i\sigma]$. For that a permutation $\pi_i$ and its inverse $\pi_i^{-1}$ are built as follows. The permutation $\pi_i$ is built as follows: $$\pi_i[j]=C_i[b[j]]+\mathtt{rank}_{b[j]}(j)$$ 
where $$C_i[c]=\sum_{1\leq k<c}{f_{c,k}}$$

and the rank queries are relative to the sequence $b_i$. In other words, the permutation $\pi_i$ is built, by replacing 
a character $c$ that occurs at position $j$ in $b_i$ by the number of occurrences of character $c$ in $b_i$ up to position $j$, 
plus the total number of occurrences of all characters $c'<c$ in $b_i$. 

The inverse permutation $\pi^{i}$ is built by enumerating the positions of character $1$ in $b_i$, followed by the positions 
of occurrences of character $2$ in $b_i$,\ldots, followed by positions of occurrences of character $\sigma$ in $b_i$. 

It is then easy to see that :
$$\pi^{-1}_i[j]=\mathtt{select}_{c}(j-C[c])$$ 

where $c$ is the only character such that $C_i[c]<j\leq C_i[c+1]$. 
It is easy to see that simulating random access to $\pi_i$ can be done by using an \textsc{access} and a partial rank operation to the sequence $b_i$. It is equally easy to see that a random access to $pi^{-1}_i$ can be done through a predecessor query on the array $C_i$ followed by a \textsc{select} operation on the sequence $b_i$. 
A technique from~\cite{munro2003succinct} allows to store (implicitly) 
only one of $\pi$ or $\pi^{-1}$ so that random access 
to any element of the permutation is supported in constant time. 
Random access to the other permutation is achieved in $O(c)$ time, provided 
that one uses space $O(f_{c,i}\log\sigma/c)$ bits of space. By choosing $c=\log\log\sigma$, 
$O(\log\log\sigma)$ query time and space $O(f_{c,i}\log\sigma/\log\log\sigma)$. 

In order to store $\pi$, we need to store a plain representation 
of the array $b_i$ in $\sigma\log\sigma$ bits. Then, in order 
to support partial $\mathtt{rank}_{b[j]}(j)$, we can use a monotone 
minimal perfect hash function on the positions 
of occurrences of character $c$ in $b_i$ 
which would use $O(f_{c,i}\log\log\sigma)$ bits of space. The mmphf 
can be built in randomized $O(f_{c,i})$ time (see section~\ref{sec:build_mmphf}). 

In order to store $\pi^{-1}$, one uses an indexed bitvector to represent the array $C$
in $2\sigma$ bits. Then a \textsc{select} query on that bitvector for a given position
$j$, will allow to determine the character $c$ such that $C_i[c]<j\leq C_i[c+1]$.
In order to support $\mathtt{select}_{c}(j-C[c])$ on $b_i$, one needs 
to use prefix-sum data structure for the occurrences of character 
$c$ which would use space $f_{c,i}(\log (\sigma/f_{c,i})+O(1))$ bits allow 
select operation in $O(1)$ time. The prefix-sum data structure can be built 
in time $O(f_{c,i})$. 

We now describe in more detail the technique 
of~\cite{munro2003succinct}. The idea is to consider 
a permutation $\pi$ of size $t$ as a collection of cycles
(it could be just one cycle or $t$ cycles). Let us 
define by $\pi^t[i]$ for $t>1$, as $\pi[\pi^{t-1}[i]]$ 
and $\pi^1[i]=\pi[i]$. 
Consider the smallest integer $t\geq 1$ such that 
$\pi^{t_i}[i]=i$, then one could say that the cycle that
contains $\pi[i]$ is of size $t$. It is easy then, to 
decompose the permutation $\pi$ into cycles by first 
building the cycle $1,\pi[1],\pi^2[1]\ldots \pi^{t_1-1}[1]=1$ 
(by iteratively applying $\pi$ until we get the value $1$) 
and then marking all the elements in the cycle. Then, one could 
build another cycle by picking the next non marked position $i>1$. 
The main idea is to break each cycle of length more than $t>c$ into 
$\lceil t/c\rceil$ blocks of length $c$. Then, store in a dictionary, 
the first element in a block and associate with it a back pointer 
to the first element in the block that precedes it. 
That is, given a rotation $x,\pi[x],\pi^2[x],\ldots,\pi^{t-1}[x]$ 
the dictionary will store the pairs of key values $(\pi^{ic+1}[x],\pi{(i-1)c+1}[x])$, 
for all $i>1$ and the pair $(x,\pi{t-c}[x])$. Then, determining $\pi^{-1}[y]$ for any value $y$
can be done in $O(c)$ time, by successively computing $y,\pi[y],\pi^{2}[y]\ldots,\pi^{c}[y]$ and querying 
the dictionary for every element in the sequence. Then the query will be successful, for some 
$\pi^{i}[y]$ with $i\in[0,c-1]$ and the dictionary will return the pair $(\pi^{i}[y],\pi^{i-c}[y])$. 
Then it suffices to compute the sequence $\pi^{i-c}[y],\pi^{i-c+1}[y],\ldots,\pi^{-1}[y]$.
If one chooses $c=\log\log\sigma$, then the space bound for storing 
the dictionary will be $O(n\log\sigma/\log\log\sigma)=o(n\log\sigma)$, the time 
to construct it is deterministic $O(n\log\sigma/\log\log\sigma)=o(n\log\sigma)$
(say representing the dictionary using bitvector of size $n$ with $n\log\sigma/\log\log\sigma$
ones) and the time to compute $\pi^{-1}[y]$ for any value $y$ is $O(\log\log\sigma)$.

\subsection{Building monotone minimal perfect hashing}
\label{sec:build_mmphf}
We prove the following lemma:
%proved in appendix~\ref{sec:build_mmphf}

\begin{lemma}
\label{lemma:build_mmphf}
Given a set $S$ of $n$ elements from universe $U$ in sorted order, we can build in $O(n)$
(randomized ) time a monotone minimal perfect hash function that occupies $O(n\log\log(U/n))$
bits of space and that answers to queries in $O(1)$ time. 
\end{lemma}
The monotone minimal perfect hash function~\cite{BBPV09} can be built in randomized $O(n)$ time on $n$ sorted elements. 
We first recall the monotone minimal perfect hash function. Given a set $S$ represented in sorted 
order by the sequence $x_1<x_2<\ldots <x_n$
where $x_i\in [1..U]$, we first partition the sequence into $\lceil n/b\rceil$ blocks
of consecutive elements where each block has $b=\log n$ elements (except possibly for the last block). 
We then for each block $i$ compute $y_i$ the longest common prefix of the elements 
in the block (starting from the most-significant bit)
of the block $x_{i(b-1)+1}\ldots x_{ib}$ ($x_{i(b-1)+1}\ldots x_n$ for the last block). 
In order to compute the length of the longest common prefix of the elements in the block, 
we can take the longest common prefix of first and last element in the block(
$x_{i(b-1)+1}$ and $\ldots x_n$ for the last block and $x_{i(b-1)+1}$ and $x_{ib}$
for the others) which can be computed
in constant time using the $MSB$ operation (which returns the most-significant bit 
of a given number) which can be simulated using constant number of multiplications~\cite{brodnik1993computation}. 

We then in linear time~\cite{HT01},build a minimal perfect hash function $F$ on the set 
$S$ which maps every key in $S$ to an index 
in $[1..n]$. We then store a table $\ell[1..n]$ of cells where $\ell[F(x_i)]=|y_{\lfloor(i-1)/b\rfloor+1}|$. 
In other words every key of a block $i$ stores the length of $y_i$ at the position $F(x_i)$ in the table $\ell$.
We also use a table $r[1..n]$, where $r[F(x_i)]=i-b\cdot\lfloor(i-1)/b\rfloor$. In other words $r$ stores 
the rank of an element in its block.
 
The total space usage of $F$ is $O(n+\log\log U)$ bits and the total space usage of $\ell$ will be $O(n\log\log n)$ bits. 

It has been shown that all $y_i$ are distinct for all $i\in[1..\lceil n/b\rceil]$, so that every $y_i$ 
uniquely represents the block $b$. We then build a minimal perfect hash function $G$ on the set 
$y_1,y_2\ldots y_{\lceil n/b\rceil}$. We finally build a table $R[1..\lceil n/b\rceil]$, 
where $R[G(y_i)]=i$. The hash function $G$ occupies $O(n/\lceil n/b\rceil+\log\log u)=O(n/\log n+)$
and the table $R$ occupies $O(n\log n/\lceil n/b\rceil)=O(n)$ bits of space. 

Given a key $x_i$ we get that $i=R[G(\mathrm{pref}(x,\ell[F(x_i)])]\cdot b+r[F(x_i)])$, where $\mathrm{pref}(x,p)$
gives the prefix of $x$ of length $p$. Given $x_i$ we compute $f_i=F(x_i)$ in constant time, 
then $\ell[f_i]=|y_i|$ and $y_i=\mathrm{pref}(x,|y_i|)$. Then $R[G(y_i)$ gives us the number of the block 
that store $x_i$ and $r[f_i]$ gives the rank of $x_i$ in its block. Thus the final rank of $x_i$ 
is given by $i=R[G(y_i)]\cdot b+r[f_i])$.

We now show how to construct minimal perfect hash functions in case 
$n$ is very close to $U$. We can show that we can achieve $O(n\log\log (U/n))$ 
bits of space. For that we cut the interval $[1..U]$ into $n'\leq n$ chunks of size $\lceil U/n\rceil$
(except for the last chunk which could possibly be of smaller size) and for each chunk 
$i\in[1..n']$ with more than on element build build a monotone minimal perfect hash $f_i$ function on the
elements that fall in the chunk. We also build a prefix-sum data structure that stores the 
number of elements in each chunk and that occupies $O(n')\leq O(n)$ bits of space.
Given a query $x$, we first find the chunk that contains $x$. 
That is the chunk $i=\frac{x-1}{\lceil U/n\rceil}+1$ and count $r_0$ the number of elements in 
$[1..i-1]$ and finally compute the rank of $x$ as $f_i(x)+r_0$. 
\subsection{Building range color reporting structure}
\label{sec:build_range_color_rep}
In this section, we will show that we can build a range color reporting data structure for 
a sequence of length $n$ over an alphabet $[1..\sigma]$ in total (randomized) time $O(n)$. 
We use the method proposed in~\cite{BNV13}. The method uses the following idea which was first proposed
in~\cite{Sa07b}. We use two \textsc{rmq}s on top of the sequence of colors. The first one 
allows to enumerate all the leftmost occurrences of all distinct colors. The second allows to enumerate
all the rightmost occurrences of the colors. We then use a monotone minimal perfect hash function built 
using Lemma~\ref{lemma:build_mmphf} to compute the rank of the leftest occurrences and the rank of the rightest occurrence of each color. 
The frequency of the color is obtained by subtracting the first rank from the second and incrementing 
the result by $1$. 
We can construct the range minimum and range maximum query data structures in linear time 
and using $O(n)$ bits of extra-space using the algorithm 
of~\cite{Fi10}. Each of the two occupies $4n+o(n)$ bits. We can build the monotone minimal 
perfect hash function in $O(n\log\sigma)$ bits of space and randomized $O(n)$ time using 
the algorithm described in previous section. 
The final space for all the minimal perfect hash functions will be $O(n\log\log\sigma)$ bits
of space. 

\section{Efficient FM-indices for large alphabets}
\label{sec:build_ext_fm_indices}

\subsection{Building the Weiner link support}
\label{sec:build_weiner_links}
We can show that we can build the data structure recently proposed in~\cite{BNtalg14} in 
(randomized) linear time and compact space. 
Given the suffix tree, we consider all the nodes whose path is prefixed by 
a character $c$. For every node $x$ whose corresponding $\mathtt{bwt}$ 
interval contains character $c$ and whose corresponding path is $p$, we know that there 
exists a path $cp$ in the suffix tree. 
We traverse the suffix tree nodes in preorder, and for each node $x$ with corresponding 
interval $\mathtt{bwt}[l_x,r_x]$ and a corresponding path $p$ use the method described 
in subsection~\ref{subsec:linear_time_bwt_interval_enum}  (Lemma~\ref{lemma:enum_intervals})
to determine all the $d$ distinct characters $c_1,\ldots c_d$
that appear in $\mathtt{bwt}[l_x,r_x]$ and for each character $c_i$ 
determine the node $y$ that is the target to a Weiner link from $x$ and labeled
by character $c_i$. This takes time $O(d)$.

%We can determine the target of the Weiner link using two 
%$\mathrm{rank}$ queries in time $O(\log\log\sigma)$ which will give us an interval 
%$[l_y,r_y]$. We then determine the node $y$ by doing $\mathrm{LCA}$ query on the leaves 
%number $l_y$ and $r_y$. 
We can now determine whether the Weiner link is explicit 
or not. That is whether the path of $y$ is $c_ip$ or not. In order to do that 
we apply a suffix link on $y$ and determine whether the target node is $x$ 
or not. If so we conclude that the Weiner link is explicit, otherwise we 
conclude it is implicit. 

We now describe the used data structures. For each character $c$, we 
maintain two vectors. A vector $V_c$ that stores all the source nodes 
of Weiner links labeled by $c$. The nodes are stored in sorted order and 
we use Elias $\delta$ or $\gamma$ encoding~\cite{El75} to encode the difference between two consecutive 
nodes. In addition we store a bit vector $b_c$ that states whether each Weiner link 
is implicit or explicit. 

The two vectors $V_c$ are filled while traversing the suffix tree. Each time 
we determine that a node $x$ is the source of Weiner link labeled with character 
$c$, we append $x$ to the vector $V_c$. Then, if the Weiner link is explicit, we 
append a $1$ to vector $B_c$. Otherwise we append a $0$. 

Given that the number of Weiner links 
will be linear, the total space used by the vectors 
$B_c$ will be $O(n)$ bits. The total space used by the vectors $B_c$ will 
be $O(n\log\sigma)$. This is by log-sum inequality. 

One detail we should take care about is that of memory allocation. 
In order to avoid fragmentation, 
we will only use static memory allocation. That is, in a first pass, we will not 
store $V_c$ or $B_c$, but just determine their sizes. We will thus just maintain two 
counters $C_{V_c}$ and $C_{B_c}$ that store the number of bits needed to store 
$V_c$ and $B_c$ in addition to a variable $\mathtt{Last}_c$ that stores the last node visited 
node with Weiner link labeled by $c$. That way while traversing every node of the suffix tree
we can easily determine for each Weiner link labeled $c$, the 
space needed to store the difference between the node (using $delta$ or $gamma$ coding)
and its predecessor in the list of Weiner links labeled by $c$. 
We note that maintaining the arrays $C_{V_c}$ and $C_{B_c}$ 
requires $O(\sigma\log n)=o(n)$ bits of space. 

Once we have built the arrays $B_c$ and $V_c$ for all $c\in[1..\sigma]$, 
we build the monotone perfect hash functions $f_c$ based on $V_c$
for all $c\in[1..\sigma]$. 
For that, we will use the scheme that was presented in the previous section. 
Recall that $V_c$ stores all the nodes which have a Weiner link 
labeled with character $c$ in sorted order. Suppose that $V_c$
has $t_c$ elements, we will cut $V_c$ into intervals of

Finally for building the RMQ data structures we use the 
result of~\cite{Fi10} which uses $O(n)$ bits of space 
in addition to the original array. The building time 
is $O(n\cdot t_{\mathtt{access}})$, where $t_{\mathtt{access}}$ is the time 
to access the $A$ array, where $A[i]=j$ is the largest index $j<i$
such that $\mathtt{bwt}[j]=\mathtt{bwt}[i]=c$, otherwise $A[i]=0$ 
if there is no such $j$ . It is easy to see that $A[i]$ 
can be recovered using $\mathtt{select}_{c}(\mathtt{rank}_{c}(i)-1)$. 
As it is too costly to use the rank operation (which cost $O(\log\log\sigma)$ 
time), we note that $A[i]$ can be obtained in constant time through the monotone 
minimal perfect hash function built on $\mathtt{bwt}$ in linear time. 

\begin{lemma}
\label{lemma:weiner_link_support}
Given a text of length $n$ over an alphabet of size $\sigma$ 
whose \textsc{bwt} and suffix tree topology 
have been precomputed, we can in $O(n)$ randomized time and using $O(n\log\sigma)$
bits build a data structure that occupies $O(n\log\log\sigma)$ bits of space
and that allow to compute a Weiner link in time $O(t_{\mathtt{select}})$, 
where $t_{\mathtt{select}}$ is the time to do a \textsc{select} query on the \textsc{bwt}. 
\end{lemma}
We thus conclude with the following theorem:
\begin{theorem}
Given a text of length $n$ over an alphabet of size $\sigma$, 
we can in randomized $O(n)$ time and $O(n\log\sigma)$ bits 
of space, build an index that occupies $n\log\sigma(1+o(1))+O(n(\log n)/d)$ 
bits of space and that can count the number of occurrences 
of a string of length $m$ in time $O(m)$ and then report occurrences of 
the pattern  in $O(d)$ time per occurrence. It can also extract an arbitrary
substring of the text in $O(d)$ time. 
\end{theorem}

Note that the data structure referred in the theorem is described in~\cite{BNtalg14}. 
For and older and slightly slower variant we can show deterministic 
construction time:
\begin{theorem}
Given a text of length $n$ over an alphabet of size $\sigma$, 
we can in deterministic $O(n)$ time and $O(n\log\sigma)$ bits 
of space, build an index that occupies 
$n\log\sigma(1+o(1))+O(n(\log n)/d)$ bits of space and that can count the number of occurrences 
of a string of length $m$ in time $O(m\log\log\sigma)$ and then report occurrences of 
the pattern  in $O(d)$ time per occurrence. It can also extract an arbitrary
substring of length $m$ of the text in $O(d+m)$ time. 
\end{theorem}

\subsection{Building the bidirectional Weiner link support}
\label{sec:build_biweiner_links}

We can show that we can build the data structure recently proposed in~\cite{BCKM13}. 
We will build a sequence of Balanced parentheses $L_c$ for each character 
$c$. That sequence will store the topology of a virtual tree that represents 
all destinations of Weiner links that are labeled by $c$. 
We will also build a vector $N_c$ that represents all the suffix tree 
nodes that are source of Weiner links labeled by $c$. 
Our goal is to fill a vector $\delta_c$ that for each node with Weiner 
link labeled by $c$, stores the difference  between the number of 
occurrences of characters $b<c$ in the associated suffix array interval and the same 
in all its closest descendant that have a Weiner link labeled by $c$. 

We assume that we have already built the unidirectional Weiner link support
(using Lemma~\ref{lemma:weiner_link_support}). 
We first traverse the suffix tree nodes in depth first order. The traversal 
does not need to use a stack. We use $\mathtt{next\_sibling}$, $\mathtt{parent}$ 
and $\mathtt{child}$ operations. Each node (except the leaves) 
is traversed twice, once in descending and in ascending 
directions. 
For each traversed node $x$ with corresponding 
interval $[l_x,r_x]$ we enumerate distinct characters $c_1,\ldots c_d$
that appear in $\mathtt{bwt}[l_x,r_x]$. That is all the characters 
that labels Weiner links starting from $x$. 
When we traverse $x$ in descending direction. For each character $c_i$, we append an 
opening parenthesis to sequence $L_c$. When we traverse $x$ in ascending 
direction, we append a closing parenthesis to $L_c$ and append the value 
of $x$ to the vector $N_c$. We do not store the value itself, but instead 
store the difference between $x$ and the last node stored in $N_c$ (we use 
a vector $Last_c$ to store the last node in $N_c$). 

Once we have built the Weiner link support for each character $c=2,\ldots ,\sigma$. 
Recall that our goal is to build for each character $c$ a data structure that stores 
for each node $x$ that has a Weiner link labeled by $c$, the number 
of occurrences of characters $b<c$ in the suffix array intervals that corresponds to node $x$. 

We will use a temporary vector $\mathtt{lastchar}[1..2n-1]$, 
where $\mathtt{lastchar}[i]$ stores the index of the last character. 
The vector occupies $O(n\log\sigma)$ bits of space. 
Initially all entries are set to zero. 
We will also use a temporary stack of capacity at most $O(n)$ bits. 
For a given character $c$, our first step will be to build a tree topology 
data structure on top of vector $L_c$. 
We then traverse the vector $N_c$ and 
for each node $x$ compute the number of occurrences of characters $b<c$ 
in the interval $\mathtt{bwt}[l,r]$ that corresponds to node $x$. This is done as follows:
we retrieve character $c'=\mathtt{lastchar}[x]$. This character 
is the largest character $c'$ such that $c'<c$. We then 
use the (unidirectional) Weiner link support to count the number of occurrences of the 
character $c'$ in $\mathtt{bwt}[l,r]$ in constant time time. 
We then use the bidirectional Weiner link support for character $c'$ (which by definition 
has already been built since $c'<c$) to count the total number 
of characters $b<c'$ in $\mathtt{bwt}[l,r]$ and add the total to the number of occurrences of 
$c'$ in $\mathtt{bwt}[l,r]$. This will give us the total number of occurrences of characters 
$b<c$ in $\mathtt{bwt}[l,r]$ (which we note by $\alpha_{c,x}$).

Whenever $x$ is a leaf we will simply push on the stack 
a $1$ if the interval $\mathtt{bwt}[l,l]$ contains a single 
character $b<c$ and push a $0$ otherwise. 

If $x$ is an internal node, we will pop from the stack the counters 
associated with all the children of $x$. That is we use the topology
built on top of $N_c$ to retrieve the number of children of $x$, 
and then sum up their value. We will then subtract the sum from 
$\alpha_{c,x}$. We then append $\alpha_{c,x}$ at the end of vector 
$\delta_c$ (we use gamma-coding to push the value).

We then push the original value of $\alpha_{c,x}$ (before subtraction) 
on the stack (actually the value is delta or gamma coded before 
pushing on the stack). We finally set $\mathtt{lastchar}[x]=c$. 

\begin{lemma}
\label{lemma:bi_weiner_link_support}
Given a text of length $n$ over and alphabet of size $\sigma$ 
such that the \textsc{bwt}, the suffix tree topology and 
the Weiner link support for both the text and its reverse 
have been precomputed. Then in $O(n)$ randomized time 
and using $O(n\log\sigma)$ bits of space, we can build a data structure 
that occupies $O(n\log\sigma)$ bits of space and that allows to compute a 
bidirectional Weiner link (and suffix link) in time $O(1)$.
\end{lemma}
We thus conclude with the following theorem:
\begin{theorem}
\label{theo:full_double_BWT}
Given a text of length $n$ over an alphabet of size $\sigma$, 
we can in randomized $O(n)$ time, build an index that occupies 
$O(n\log\sigma)$ bits of space and that support operations $\mathtt{extendleft}$, 
$\mathtt{extendright}$, $\mathtt{isLeftMaximal}$, $\mathtt{isRightMaximal}$, 
$\mathtt{enumerateLeft}$ and $\mathtt{enumerateRight}$ in $O(1)$ time. 
\end{theorem}
Note that the data structure built in Theorem~\ref{theo:full_double_BWT} 
is precisely the data structure number $3$ described in~\cite{BCKM13}. 

Also, by combining lemmata~\ref{lemma:weiner_link_support},~\ref{lemma:bi_weiner_link_support} and~\ref{lemma:plcp} 
we get the following lemma:
\begin{lemma}
\label{lemma:plcp2}
Given a text of length $n$ over an alphabet of size $\sigma$ 
whose \textsc{bwt} and the suffix tree topology 
and the \textsc{bwt} of its reverse have been precomputed,
we can build the \textsc{plcp} array in randomized 
$O(n)$ time and $O(n\log\sigma)$ bits of 
additional space. 
\end{lemma}
\section{Other compressed suffix tree operations}
~\label{sec:build_cst}
Most of the suffix tree operations but not all can be supported in constant time by using combination of the suffix tree topology, the permuted lcp array and the FM-index. In particular, three important operations are supported in time $O(\log^\epsilon n)$. These are the string depth operation, the child operation and the String level ancestor operation. The first operation is supported in time $O(t_{\mathtt{SA}})$ using all three components of the CST (which were previously shown). We show how that auxiliary data structures necessary to support the last two operations can be built in time randomized $O(n)$ time. 

\subsection{Suffix tree with blind child support}
\label{sec:build_blind_child_support}
In~\cite{BN11,BNtalg14} it was shown how to augment a suffix tree with $O(n\log\log\sigma)$
bits of space so that a child operation can be supported in $O(t_{\mathtt{SA}})$ (which translates 
to $O(\log^\epsilon n)$ time for the \textsc{csa} version that uses $O(n\log\sigma)$ bits of space). 
We can show that the augmentation can be built in $O(n)$ time as follows. 

Once we have built the suffix tree topology, we use Lemma~\ref{lemma:enum_intervals} to 
generate all suffix array intervals that correspond to internal suffix tree node along with
their child labels (in sorted order), and for each of interval
determine the suffix tree node using the tree topology. 
We finally store the labels of all children of the node 
in an $A$ array of total size at most $(2n-2)\log\sigma$ bits, which stores all the children labels
of all nodes, where the nodes are sorted in order. That is for every node $x$, we store 
the child labels in positions $A[f(x)..f(x)+g(x)-1]$, where $f(x)$ counts the total number of children of all 
nodes $x'<x$ and $g(x)$ counts the number of children of node $x$. We then scan the array $A$ and 
build the monotone minimal perfect hash function on the child labels of every node $x$. 

\subsection{String level ancestor queries}
\label{sec:build_blind_child_support}

The string level ancestor is an important operation on the suffix tree. It can be supported in time $O(t_{\mathtt{SA}}\log\log n)$~\cite{KKNS13} (which translates to $(\log^\epsilon n)$ time using the \textsc{csa} version that uses $O(n\log\sigma)$ bits of space). 
We now show that the support for string level ancestor queries can be added in time $O(n)$ on top of a compressed suffix tree representation. The additional space is $o(n)$. 
We first describe the string level ancestor operation implementation as explained to us by Travis Gagie~\footnote{We thank Travis Gagie for explaining it to us.}. This implementation is different from the one described in described in~\cite{KKNS13} but achieves essentially the same time and space bounds.

We sample every $b=\log^2n$ node in the suffix tree and build a weighted level ancestor (\textsc{wla} for short) data structure  on it~\cite{FM96}, where the weight associated with every node will be its string depth. 
We now describe how the sampling is done. We first define the height and the depth of the tree nodes. We define the depth
of a node as the distance between the root and the node. We define the height of an internal node $N$ as the difference between the depth of the deepest leaf in the subtree rooted at $N$ and the depth of $N$. 
A node will be sampled if and  only if:
\begin{enumerate}
\item Its depth is multiple of $b$. 
\item Its height is at least $b-1$. 
\end{enumerate}
We can easily show that the number of sampled nodes will be at most $n/b$. This is easy to see. For every sampled node, we can associate at least $b-1$ non-sampled nodes. Suppose that the sampled node $N$ has no sampled node among its descendants. Then $N$ has height at least $b-1$ and thus must have a path that contains at least $b-1$ non-sampled nodes and we can associate all the nodes in that path with $N$. Otherwise, the sampled node $N$ is at depth $ib$ and has at least one descendant at depth $(i+1)b$ and all the $b-1$ nodes along the path between the two sampled nodes will not be sampled. Thus, we associate all those $b-1$ nodes with $N$. 

We now describe the sampling algorithm. It can be done using navigation operation $\mathtt{first\_child}$, $\mathtt{next\_sibling}$ and $\mathtt{parent}$ and operations $\mathtt{depth}$ and $\mathtt{height}$. We do a full traversal of the suffix tree and for every node $N$ of depth $d$ multiple of $b$ and height at least $d+b-1$, we determine the string depth $\mathtt{SD}_N$ of $N$ and append the pair $(N,\mathtt{SD}_N)$ to a list initially empty. During the traversal, we can easily generate the tree topology of the tree that contains only sampled nodes by generating a sequence of balanced parenthesis in the following way:
every time we traverse down a sampled node, we append an opening parenthesis and every time we traverse up we append a closing parentheses. We then generate a dictionary $D$ that stores all the sampled nodes. This would use $O((n/b)\log n)=O(n/\log n)=o(n)$ bits. 
At the end we use the list of pairs $(N,\mathtt{SD}_N)$ and the tree topology as input to an algorithm that generates the \textsc{wla} data structure~\cite{FM96}. 

We now describe how queries are implemented. We first note that the level ancestor query which asks given a node $N$, to return the ancestor at depth $d$ can be supported in constant time using the representation of~\cite{SNsoda10}. 

Given a node $N$ and a target string depth $\mathit{sd}$, we first determine the depth $d$ of $N$. We then make a level ancestor query to determine the node $M$ at depth $ib$, where $i=\lfloor d/b\rfloor$. We then query $D$ to find whether the node $M$ is sampled or not. If $M$ is not sampled and is not the root, then we replace $M$ by its ancestor at depth $(i-1)b$. If $M$ is the root, then its string depth is $0$ and we keep it. 

We then query the \textsc{wla} data structure to determine the node $L$, the deepest sampled node ancestor of $M$ and whose depth is at most $\mathit{sd}$ (we return $M$ if its depth is at most $\mathit{sd}$). If that node is $M$ itself, then we need to do a binary search over the ancestors of $L$ of depths between $ib$ (or $(i-1)b$ if $M$ has been replaced) and $d$ by using the string depth and the level ancestor operations on the compressed suffix tree. If $L\neq M$ and $L$ is at depth $jb$. Then we need to do a binary search over ancestors of $N$ of depths between $jb$ and $(j+1)b-1$. 

The total time is dominated by the binary search over a depth interval of size at most $2b$ which takes $O(\log b)=O(\log\log n)$ steps and time $O(\log\log n\cdot t_{\mathtt{SA}})$.
\subsection{Completing the compressed suffix tree construction}
Theorem~\ref{theo:cst_large_alphabet} is proved by combining Theorem~\ref{theo:build_bwt} with lemmata~\ref{lemma:tree_topology2} and~\ref{lemma:plcp2} and using the augmentations presented in the last two subsections. 

\section{Deterministic constructions}
~\label{sec:det_building}
We can show that the deterministic construction can be done in $O(n)$ at the price 
of slowing down the Weiner links (in the FM-indices and the compressed suffix tree) to $O(\log\log\sigma)$ time and the child operation (in the suffix tree) to $O(\log\log\sigma \cdot t_{\mathtt{SA}})$ time. For that purpose we completely eliminate 
any use of monotone minimal perfect hash functions, and instead rely on $\mathtt{rank}$ 
queries which are answered in time $O(\log\log\sigma)$. This slows down Weiner links to $O(\log\log\sigma)$
time. 
For the child operation, we notice that every node with $d$ children, we can sort the $d$ labels
and sample one in every $\log\sigma$ child and store the sampled labels in a predecessor 
structure which answers in time $O(\log\log\sigma)$. The predecessor structure can be build in $O(d)$
time and occupies $O(d)$ bits of space. The child operation is finished by doing binary search on 
an interval of $\log\sigma$ labels in time $O(\log\log\sigma\cdot t_{\mathtt{SA}})$.

\end{document}